\documentclass[aip,
jcp,
amsmath,
amssymb,
reprint, 
citeautoscript,
]{revtex4-2}

\usepackage{graphicx}

\usepackage[caption=false,labelformat=simple]{subfig}

\usepackage{dcolumn}
\usepackage{hyperref}

\begin{document}

\preprint{}

\title[
Simple and efficient methods for local structural analysis in polydisperse hard disk systems
]{
Simple and efficient methods for local structural analysis in polydisperse hard disk systems
}

\author{Daigo Mugita}
\affiliation{Graduate School of Engineering, Nagoya Institute of Technology,
Nagoya, 466-8555, Japan}

\author{Kazuyoshi Souno}
\affiliation{Graduate School of Engineering, Nagoya Institute of Technology,
Nagoya, 466-8555, Japan}

\author{Hiroaki Koyama}
\affiliation{Graduate School of Engineering, Nagoya Institute of Technology,
Nagoya, 466-8555, Japan}

\author{Taisei Nakamura}
\affiliation{Graduate School of Engineering, Nagoya Institute of Technology,
Nagoya, 466-8555, Japan}

\author{Masaharu Isobe}
\email{isobe@nitech.ac.jp}
\affiliation{Graduate School of Engineering, Nagoya Institute of Technology,
Nagoya, 466-8555, Japan}

\date{\today}

\begin{abstract}
In nonequilibrium statistical physics, quantifying the nearest (and higher-order) neighbors and  free volumes of particles  in many-body systems is crucial to elucidating the origin of macroscopic collective phenomena, such as glass/granular jamming transitions and various aspects of the behavior of active matter. However, conventional techniques (based on a fixed-distance cutoff or the Voronoi construction) have mainly been applied to equilibrated, homogeneous, and monodisperse particle systems.
In this paper, we implement simple and efficient methods for local structure analysis in  nonequilibrium, inhomogeneous, and polydisperse hard disk systems. We show how these novel methods can overcome the difficulties encountered by conventional techniques, as well as demonstrating some applications.
\end{abstract}


\maketitle

\section{\label{sec:1} Introduction}
In many-particle (molecular) systems, the use of a local configuration of neighbors and the free volume around a tagged particle constructed from the    excluded volume due to the presence of neighbors has made a crucial contribution to descriptions of macroscopic properties, elucidating their microscopic origin, and predicting dynamical behavior in many areas of physical science, such as in the theory of the liquid state\cite{krauth_2006, isobe_2016a, chandler_1987, hansen_2006}.
In recent years, much research activity has been focused on the properties of  nonequilibrium, inhomogeneous, and polydisperse systems, such as  granular (jamming) systems, systems undergoing glass transitions, and active matter\cite{maiti_2013, maiti_2014, schindler_2015, isobe_2016a, isobe_2016b, chen_2023, ballerini_2008, chandler_2010, berthier_2011, royall_2015, biroli_2013}, the understanding of which requires precise knowledge of microscopic dynamics and the local configurations of neighbors, including higher-order neighbors.
In the study of supercooled glass-forming liquids, the exploration of slow microscopic relaxation mechanisms has been a longstanding focus of research, yet it remains an unresolved challenge\cite{chandler_2010, berthier_2011, royall_2015, biroli_2013, stillinger_2013}.
Theoretical perspectives have highlighted the significance of the free volume in understanding the dynamics of glasses, particularly with regard to voids that represent empty space on the scale of particle sizes\cite{turnbull_1961}.
A recent critical advance has been the identification of quasi-voids~\cite{lam_2017} that drive string-like hopping motions in colloid experiments\cite{yip_2020}.
Although this finding holds promise as a key to unraveling various phenomena associated with glasses, current analyses are limited to the immediate vicinity of a few neighbors' shells.
There is a pressing need for a suitable methodology to understand the cooperative dynamic facilitation~\cite{isobe_2016b, keys_2011, chandler_2010} observed among a multitude of different particles at the  molecular level, a topic that has received limited  research attention to date.

Finding a suitable definition for identifying nearest neighbors (NNs) around a tagged particle in a many-particle system is one of the crucial tasks in the quantitative characterization of phase behavior. 
Phases are often evaluated in terms of local order parameters (such as the bond orientational order parameter\cite{steinhardt_1983, nelson_2002}).
The conventional schemes for determining NNs include that based on a fixed-distance cutoff and the Voronoi construction\cite{voronoi_1908}.
Many extensions and other definitions have also been used, as, for instance, in Refs.~\onlinecite{isobe_2012, gellatly_1982, medvedev_1994}.
The order parameters depend on the number of NNs and their local configuration; however, there are ambiguities in the definitions that are used to determine NNs\cite{meel_2012}. 
Furthermore, the conventional schemes encounter difficulties when applied to systems containing particles with different radii.
The introduction of  multiple parameters for the cutoff length and the use of a radical plane construction in an extension of the Voronoi construction (with no regions of volume left unallocated) have been proposed, and the latter approach has been successfully used\cite{gellatly_1982, maiti_2013, maiti_2014} to deal with some problems in polydisperse systems.
In addition, a simple algorithm to identify NNs, the solid-angle-based nearest-neighbor  (SANN) algorithm \cite{meel_2012}, has been proposed and has a number of advantages over conventional methods.
The determination of higher-order neighbors (e.g., second and third neighbors) in  monodisperse dense liquid systems using  extended methods based on a fixed-distance cutoff faces similar difficulties \cite{isobe_2012}, 
and to overcome these 
an extended definition of higher-order neighbors based on SANN has been proposed in Ref.~\onlinecite{meel_2012}. However, it is noted that this extension does not work well.

Model systems using hard spheres or disks are often used owing to their simplicity and the fact that they exhibit a  well-defined excluded volume effect\cite{krauth_2006, isobe_2016a}.
The excluded volume is defined as the region inaccessible to the center of a tagged particle because of the presence of a solid core of other particles. 
In theoretical descriptions of the liquid state \cite{chandler_1987, hansen_2006}, the radial distribution function and the free volume of a tagged particle play very important roles, and they are usually  constructed under the assumption that the system is an equilibrated homogeneous, monodisperse particle system.
In a dense glassy system, the excluded volume effect becomes dominant. 
In particular, in the glass-forming bidisperse hard disk systems that are often used as models of the glass transition~\cite{isobe_2016b}, it is important to be able to identify the precursors of the particle-sized voids that drive hopping motions~\cite{yip_2020}.

It is therefore crucial to be able to quantify  the free volume efficiently and rigorously. 
In conventional schemes for the estimation of the free volume, simple Monte Carlo sampling is adopted as an approximation method, but this suffers from  technical difficulties and also requires excessive computational resources to achieve convergence.
An essential task when estimating the free volume is to identify the neighboring particles that define its surface or curve by considering excluded volume circles.
Some rigorous estimation methods for the free volume have been proposed\cite{hoover_1979, tanemura_1983, rintoul_1995, sastry_1997}.
Since the  strategy of using the space-filled Voronoi (polyhedron) construction \cite{sastry_1997, sastry_1998} cannot be applied straightforwardly to  polydisperse (glassy) systems, it has been extended by incorporating the radical plane construction\cite{gellatly_1982}.
The associated algorithm is akin to the original Voronoi construction algorithm, and does not incur any substantial increase in computational cost compared with the latter\cite{nishio_2022}.
However, it is important to note that the Voronoi construction algorithm itself is already computationally expensive.
Recently,  cavity averages for hard spheres in the case of polydispersity and incomplete data have also been implemented\cite{schindler_2015}.

However, for the analysis of local structures  in in non-equilibrium, inhomogeneous, and polydisperse  systems, conventional methods encounter severe difficulties and their applicability is  limited.
Our main purpose in the present work is to propose simple and efficient methodologies for local molecular analyses in such systems.
We focus on the aforementioned SANN algorithm \cite{meel_2012, isobe_2016a} and clarify the difference between the results obtained using different methods, including quantitative results from  higher-order neighbor estimators  in two dimensions and for mono- and bidisperse hard disk systems.
As a rigorous algorithm for estimating the free volume in a polydisperse hard disk system, we propose a simple, efficient, and precise method of categorizing neighbors for enclosing the local free area (NELF-A), which focuses solely on the geometry of intersections between excluded volume circles, extends the pioneering concept described in Ref.~\onlinecite{hoover_1979}.

The remainder of the paper is organized as follows:
In Sec.~\ref{sec:2-0}, we revisit the conventional schemes for detecting NNs (including higher-order neighbors) and estimating  the free volume using the excluded volume.
We summarize the difficulties encountered by conventional schemes in the case of a bidisperse  hard disk system (as a representative of polydisperse systems) and describe the details of our novel implementation.
In Sec.~\ref{sec:3-0}, the efficiency, validity, and advantages of the new approach are  comparing with those of  conventional approaches. Additionally,  as an application of the novel free volume estimator, we discuss the characteristics of pressure calculation using free volume. 
Concluding remarks are given in Sec.~\ref{sec:4-0}.

\section{Numerical schemes for local configuration analysis}
\label{sec:2-0}
The methodology for analyzing the local structure of a molecular system is summarized, in particular, the methods for detecting NNs and estimating the free volume.

\subsection{Nearest neighbor estimators}
\label{sec:2-1}
In  many-body systems, to evaluate the local structure for universal assessment purposes, such as quantitatively estimation of  bond-order parameters, it would be helpful to be able to identify the NNs of a tagged particle via a unified and clear definition. 
However, such an exact definition does not exist\cite{meel_2012}. 
Instead, NNs are most commonly identified using  the fixed-distance cutoff and  Voronoi construction algorithms.
In polydisperse and heterogeneous distributed particle systems, the use of a cutoff becomes more complicated, because  dozens of parameters are required to define the cutoff radius.
With the Voronoi construction, there is a  problem of ambiguity in detecting NNs even in the case of particles located  a long distance apart when they share a very short edge of the Voronoi construction, which often causes instability of NNs.
To solve these problems arising in conventional schemes, a modern, efficient parameter-free algorithm   has been proposed \cite{meel_2012}.
In this SANN algorithm, a solid angle is assigned to each possible neighbor in three dimensions, and  the cutoff radius is then determined by imposing the requirement that the sum of the solid angles is $4\pi$.
In this section, we briefly revisit neighbor estimators, focusing on 2D systems.

\subsubsection{Fixed-distance cutoff method}
\label{sec:2-1-1}

Suppose particles interact via a short-range repulsive potential (e.g., a hard potential) in a monodisperse system.
In that case, a fixed-distance cutoff might be a better choice owing to its simplicity, since each NN and its neighbor are determined by a single parameter, namely, the cutoff distance $r_{\mathrm{c}}$.
The cutoff distance is fixed for a whole system as a global parameter to characterize  neighbors. 
Under the assumption that each particle has the same ensemble in a homogeneous equilibrated system, the cutoff distance is well described by the first minimum of the sole radial distribution function (RDF), which is related to the neighbors in the first and second coordination shells\cite{chandler_1987}.
The RDF $g(r)$ is given by the following equation in two dimensions:
\begin{equation}
g(r) = \frac{\langle n_{\mathrm{s}}(r) \rangle}{2 \pi r \rho \,dr}.
\label{eqn:rdf}
\end{equation}
Here, $\langle n_{\mathrm{s}}(r) \rangle$ denotes the system-wide average of $n_{\mathrm{s}}(r)$, where $n_{\mathrm{s}}(r)$ is the number of particles within a shell of infinitesimal thickness $dr$ at a relative distance $r$ from a tagged particle.
$\rho$ is the number density for the entire system.

\begin{figure}[!t]
\includegraphics[scale=0.1]{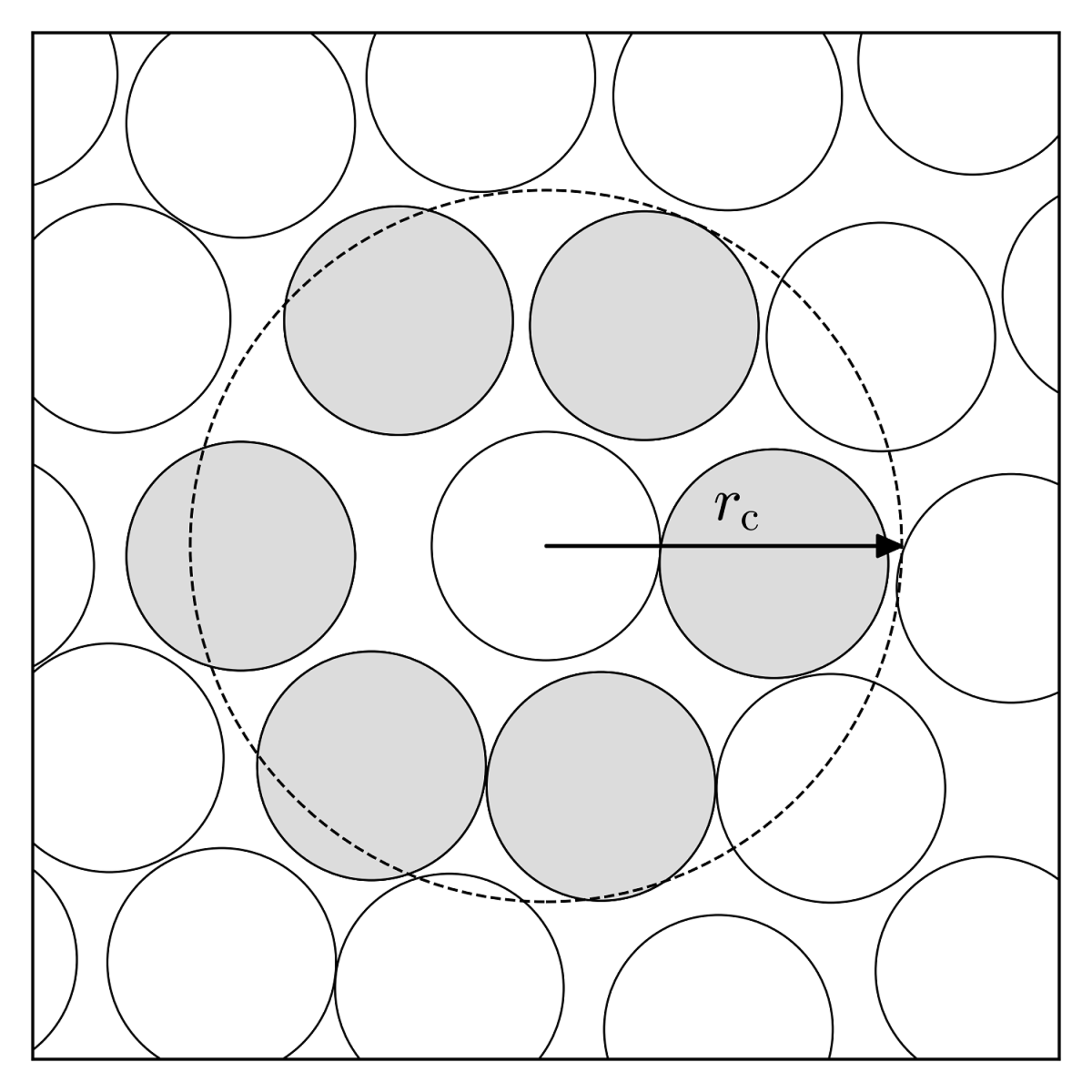}
\caption{
Schematic  of the fixed-distance cutoff in a hard disk system.
Disks inside the dotted black circle of radius $r_{\mathrm{c}}$ drawn around the center of the tagged disk  are categorized as NNs (gray-shaded circles).
}
\label{fig:cutoff}
\end{figure}

Figure~\ref{fig:cutoff} shows a typical example of the fixed-distance cutoff in a dense monodisperse hard disk system, with a radius $\sigma$ at a packing fraction $\nu= N \pi \sigma^2/ A = 0.720$, where $N$ and $A$ are the number of disks and the area of the system, respectively.
Disks whose centers are located within a length equal to the cutoff distance from the tagged disk, i.e., with $r \leq r_{\mathrm{c}}$, are detected as NNs (shown as gray-filled circles). 

\paragraph*{Difficulty}
Although this method is simple, a reasonable choice of $r_{\mathrm{c}}$ requires a preliminary calculation to estimate the first minimum of the RDF $g(r)$, which depends on the density and other details of the system.
Furthermore, since  $r_{\mathrm{c}}$ is a global parameter defined for the entire system, it becomes complex and indeed sometimes impossible to evaluate a well-defined $r_{\mathrm{c}}$ if the system has large density gradients or is polydisperse.
This situation often occurs naturally in  the study of nucleation and in glassy/jamming systems.
The first minimum of $g(r)$ is then split into several peaks due to the occurrence of different radii in such systems.
In addition, the first minimum  of $g(r)$ is often not clearly defined when there is a density gradient  induced by an external field such as gravity or the presence of an interface between two coexisting phases.

\begin{figure}[!t]
\includegraphics[scale=0.11]{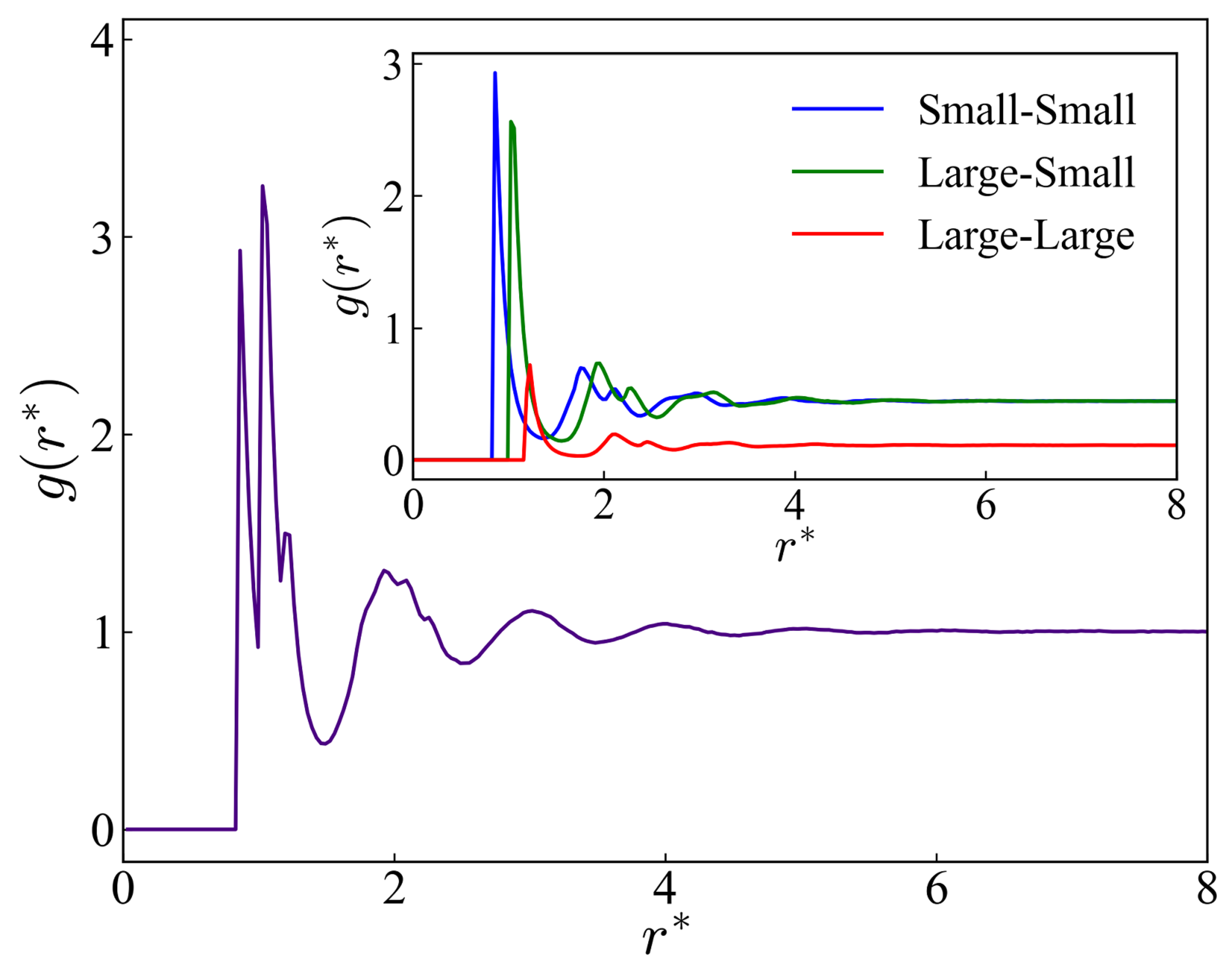}
\caption{RDFs in a bidisperse system, where $r^{\ast}=r/(2\sigma_{\mathrm{e}})$. The effective radius $\sigma_{\mathrm{e}}$ is  explained in Sec.~\ref{sec:3-3-1}. The inset indicates that the RDF can be decomposed into three parts, each corresponding to different pairings of disk size: large and large, large and small, and small and small.
}
\label{fig:rdf-bi}
\end{figure}

A modified cutoff can be considered for  polydisperse systems such as the bidisperse hard disks often used in modeling glassy systems. Figure~\ref{fig:rdf-bi} shows the RDF of a bidisperse hard disk system (the parameters are described in Sec.~\ref{sec:3-1}).  
We find that the first peak of this system is split into three different peaks, since there are three pairs of disk sizes, namely, large and large, large and small, and small and small.
It is clearly difficult to identify the first minimum of  the RDF and thereby evaluate a single global cutoff distance. 
The inset of Fig.~\ref{fig:rdf-bi} indicates that the whole RDF can be decomposed into three contributions of pairs. 
Thus, the fixed-distance cutoff can be refined by incorporating specific cutoff parameters corresponding to each pair.
By establishing three total cutoff distances, derived from the RDF for each  pair, it becomes possible to identify the NNs that encapsulate the local particle information surrounding a tagged disk.

\begin{figure}[!t] 
\includegraphics[scale=0.1]{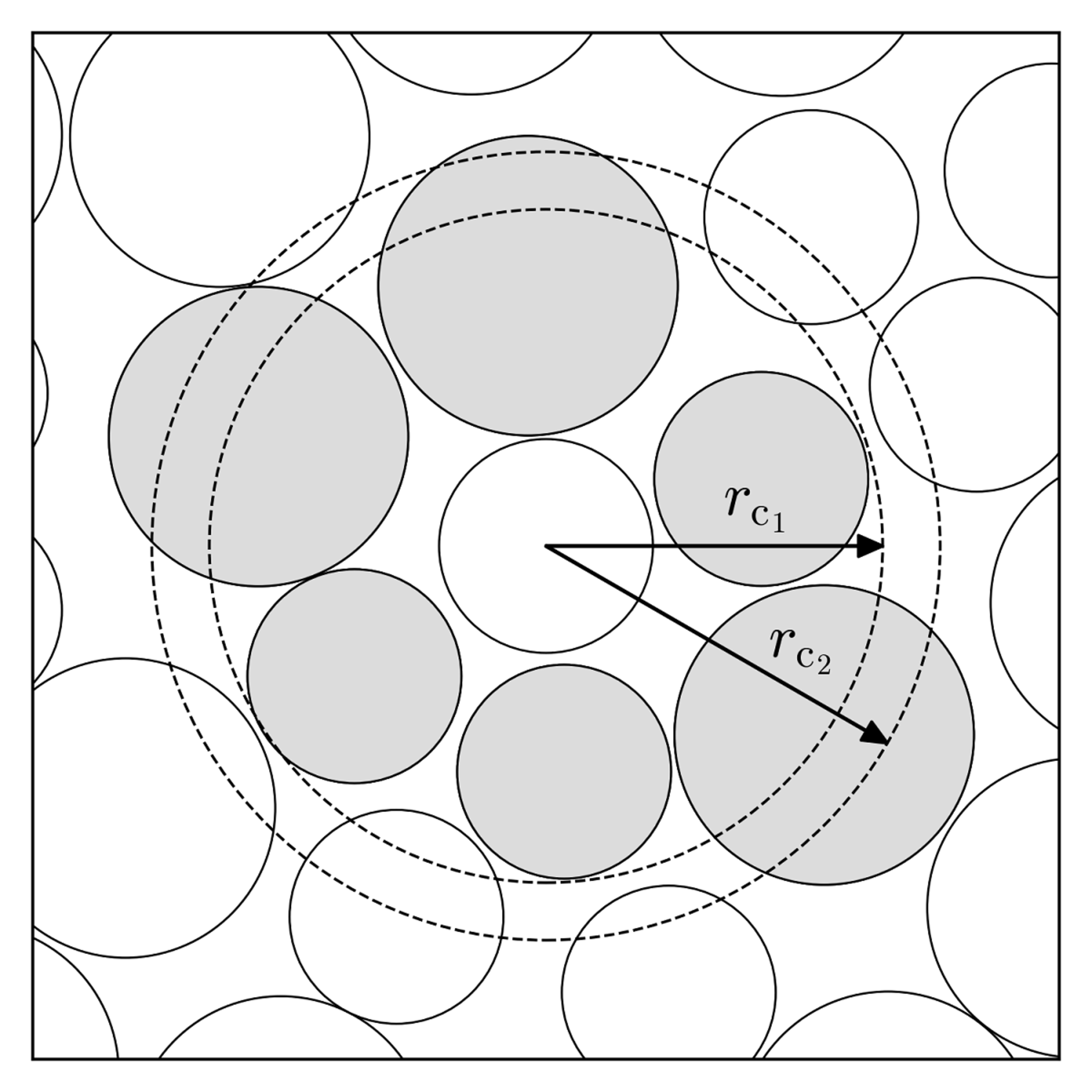}
\caption{NN distribution identified by the modified cutoff.}
\label{fig:cutoff-mod}
\end{figure}
Figure~\ref{fig:cutoff-mod} illustrates the NN distribution around a tagged disk in a bidisperse hard disk system, determined using the modified cutoff. 
For a tagged disk, two distinct cutoff radii, $r_{\mathrm{c}_1}$ and $r_{\mathrm{c}_2}$, can be independently set for small and large neighboring disks, respectively.
Three small disks are inside the circle of radius $r_{\mathrm{c}_1}$, and three large disks are inside the circle of radius $r_{\mathrm{c}_2}$.
Consequently, six disks in total are recognized as NNs. 
In polydisperse systems, as the variety of radius pairs increases, so does the  number of configurations required for accurate computation of each $g(r)$, significantly complicating the estimation of neighbors.

\subsubsection{Voronoi construction}
\label{sec:2-1-2}
A Voronoi construction for analyzing the local molecular structure is often used, simply constructed from geometric configurations with no adjustable parameter.
Those disks in the polygons in the Voronoi construction that share the edges (polyhedral faces in 3D systems) of the polygons constructed from the perpendicular bisectors are defined as NNs.
This method not only identifies the NN disks, but also provides other geometric characteristics, such as the edges, vertices, and faces between the Voronoi cells of the NNs.
Such information can be useful for local structure analysis, such as that of defects, and for phase classification.

\begin{figure}[!t]
(a)
\includegraphics[scale=0.069]{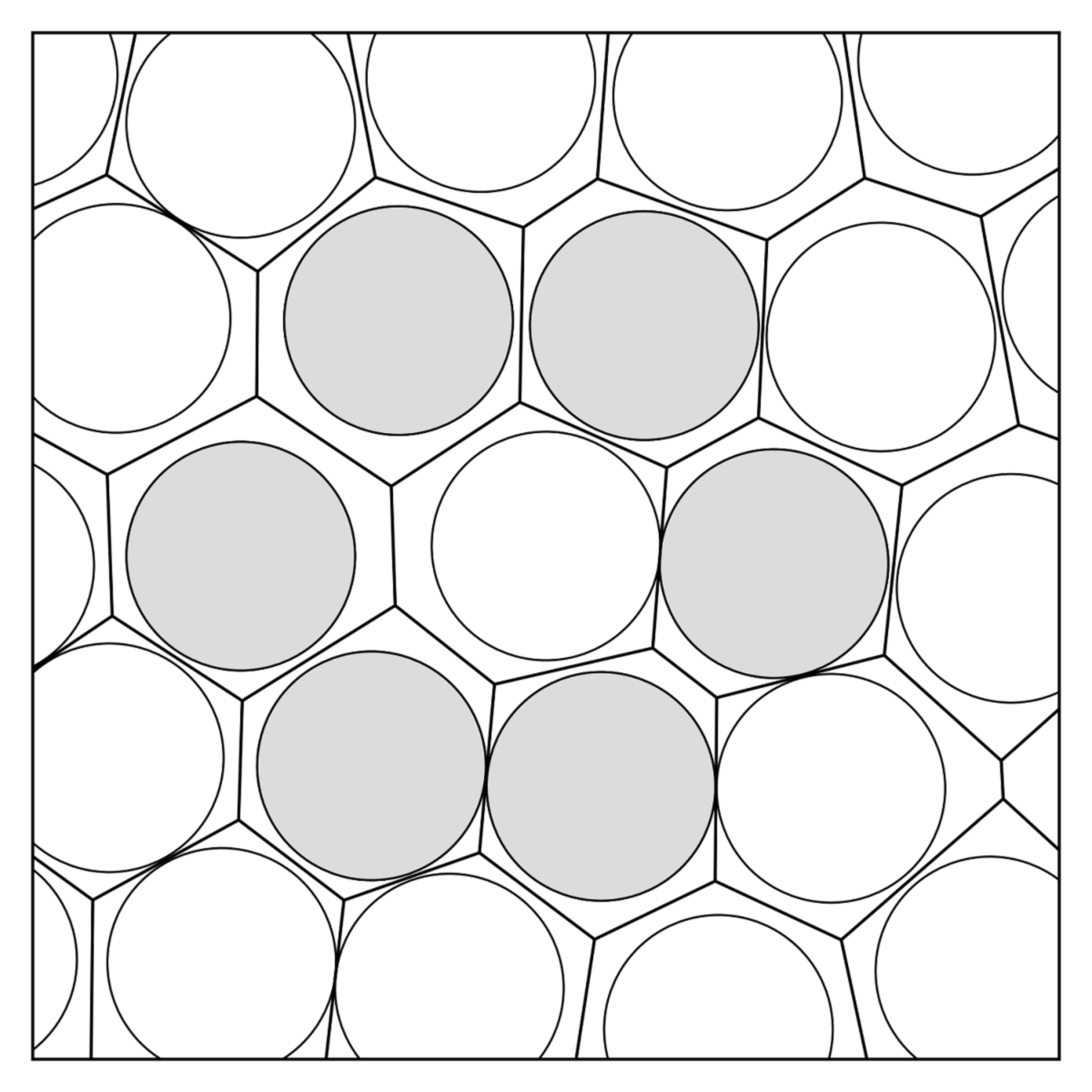}
\hspace{2mm}
(b)\includegraphics[scale=0.069]{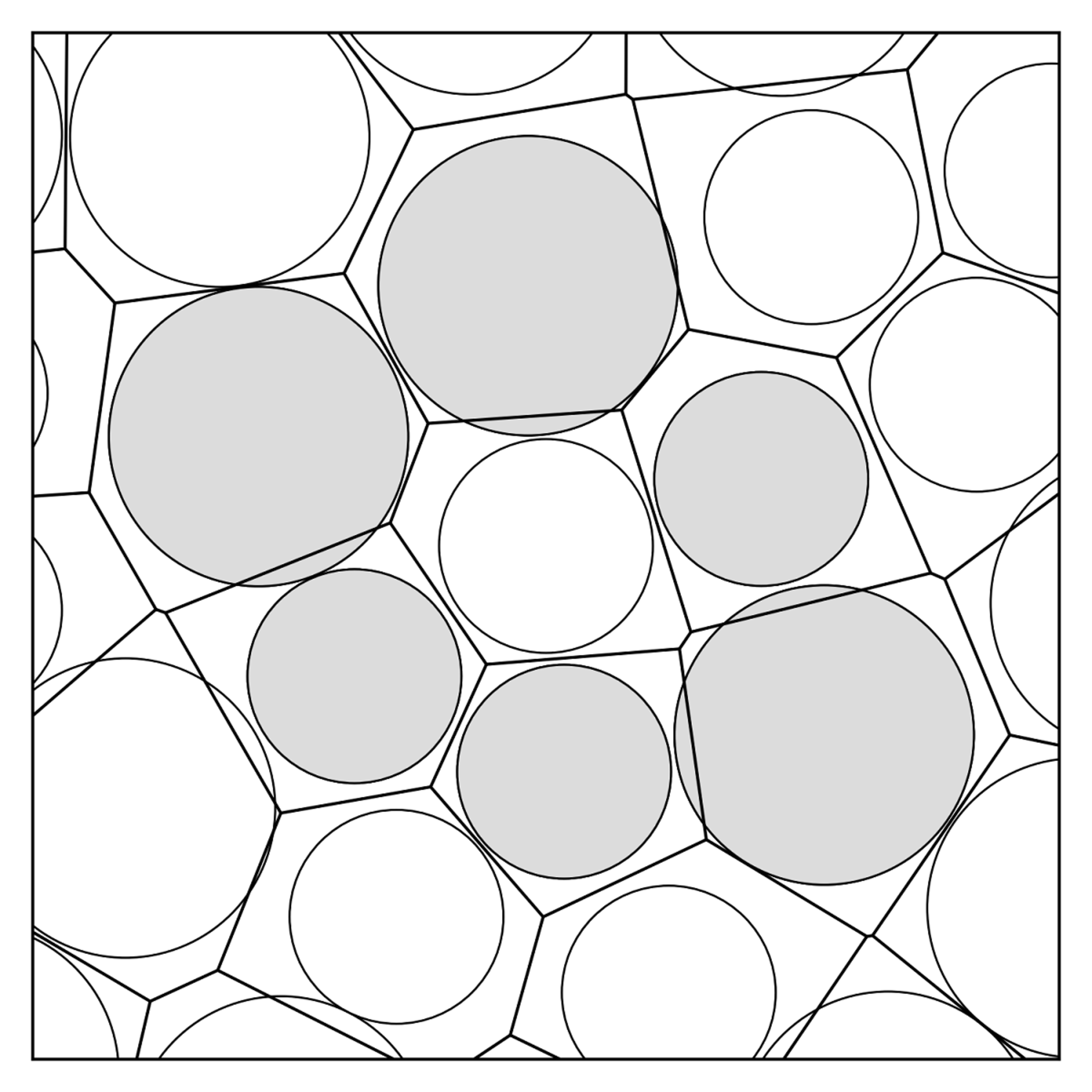}
\caption{Schematic  of  Voronoi construction in (a) monodisperse and (b) bidisperse hard disk systems. 
}
\label{fig:voronoi_mono_bi}
\end{figure}

Figures~\ref{fig:voronoi_mono_bi}(a) and~\ref{fig:voronoi_mono_bi}(b) show schematics  of the configurations around a tagged disk and its neighbors detected by the Voronoi construction (shared edges between polygons) in monodisperse and bidisperse hard disk systems, respectively. 
Six NNs are detected around the tagged disk  in both cases.
Since the Voronoi cells are constructed only on the local environment around a tagged disk, it is more applicable than the cutoff, even for inhomogeneous systems with large density gradients. 

\paragraph*{Difficulty}
The computational cost of the Voronoi construction is relatively high, and the construction is sensitive to thermal fluctuations of particle configurations.
If the particles are in a crystal structure, i.e., rattling on a lattice within a cage, they fluctuate around their equilibrium positions.
This rattling motion often causes the sharing of  tiny edges (or faces in 3D systems) with distant neighboring particles in second-coordinate shells, resulting in fluctuations and an increase in the number of NNs.  
To avoid this instability, one solution is to consider the inherent structure\cite{stillinger_1984} at a given configuration for which the local free energy is a minimum before making the Voronoi construction.
However, such an implementation, while providing robustness against thermal fluctuations, introduces specific parameters for each system and thus  loses one advantage of the parameter-free Voronoi construction.
Polydisperse systems present a challenge for precise analysis, since merely generating Voronoi polygons based on bisectors with particle centers acting as generators proves to be insufficient. For a detailed analysis in the case of polydisperse systems, it becomes necessary to incorporate an additional step that involves weighting each particle.

Weighted Voronoi tessellations extended by using the radical plane construction\cite{gellatly_1982} and Voronoi S regions\cite{medvedev_1994} are among   the solutions proposed to generate polygons for particles of different sizes.
The algorithm incorporating Voronoi S regions is more complex and computationally more expensive compared with the original Voronoi construction. By contrast, the radical plane construction requires only minor modifications to the original Voronoi algorithm and incurs almost no additional computational cost\cite{nishio_2022}.
For the purpose of identifying the index number of NNs, it is not necessary to quantify  geometric characteristics such as the edges and vertices of each Voronoi cell.

\subsubsection{SANN method}
\label{sec:2-1-3}

The SANN algorithm\cite{meel_2012} has the following advantages  compared with  conventional schemes: (i) it can be applied to systems with inhomogeneous density; (ii) it is stable against thermal fluctuations; (iii) it is parameter-free;  (iv) it is computationally inexpensive.
A 2D version of  SANN  can be applied following  the same procedure, except that it is then necessary to solve a  nonlinear equation\cite {isobe_2016a}.
The strategy adopted by  SANN  for detecting NNs is  to be simple and well-defined, and it is parameter-free even for systems with polydispersity and/or density inhomogeneities.
Additionally,  SANN  exhibits asymmetry in the sense that a given particle pair may not necessarily be identified as mutual NNs. 
This characteristic becomes pronounced in polydisperse and heterogeneous systems, making  SANN  unique compared with  other methods.
Further investigation of this property is anticipated as a topic of future research.
 In Sec.~\ref{sec:3-0}, we perform  calculations for bidisperse hard disk systems (using in particular the simple glassy model for molecular systems\cite{isobe_2016b}) to clarify the differences between the various methods.

A summary of the original concept of the SANN algorithm as presented in Ref.~\onlinecite{meel_2012} is given here. 
The relative distances $r_{i,j} =|\mathbf{r}_j-\mathbf{r}_i|$ between a tagged particle $i$ and its neighbors $\{j\}$ are sorted as $r_{i,j} < r_{i, j+1}$. 
For each particle $i$, an individual cutoff radius $R_ i^{(m)}$, called the shell radius, is introduced,  which includes the $m$ NNs of particle $i$, with $r_{i,m} \le R_i^{(m)} < r_{i,m+1}$. With each of the particles $\{j\}$ surrounding $i$ is associated an angle $\theta_{ij}$ determined by its distance from the tagged particle and the shell radius: $\theta_{ij} =\cos^{-1}(r_{i,j}/R_i^{(m)})$. Then, according to  SANN, the NNs  consist of the closest $m$ particles $\{j\}$ for which the sum of their solid angles associated with $\theta_{ij}$ equals $4\pi$, i.e.,
\begin{equation}
4\pi =\sum_{j=1}^m 2\pi(1-\cos\theta_{i,j}) =\sum_{j=1}^m 2\pi\!\left(1- \frac{r_{i,j}}{R_ i^{(m)}}\right).
\label{eqn:SANN3D}
\end{equation}
The number of NNs, $m$,  is determined from the following inequality:
\begin{equation} 
R_ i^{(m)} = \frac{1}{m-2}\sum_{j=1}^m r_{i,j} < r_{i,m+1}.
\label{eqn:SANN3Dev}
\end{equation}
To solve this inequality numerically, we start from the smallest number of NNs (i.e., $m = 3$). Then, we increase $m$ and check the inequality iteratively.
During the iterations, if the inequality~(\ref{eqn:SANN3Dev}) is satisfied at a certain $m$, we then identify that $m$ as the number of NNs $\{j\}$ that are  within the  shell  radius $R_i^{(m)}$.

For 2D systems, the original SANN algorithm can be applied following the same procedure as for 3D systems. 
The basic equation corresponding to Eq.~(\ref{eqn:SANN3D}) can be derived as follows:
\begin{equation}
2\pi = \sum_{j=1}^m 2\cos^{-1}\! \left( \frac{r_{i,j}}{R_ i^{(m)}} \right).
\label{eqn:SANN2D}
\end{equation} 
However, there is now no explicit inequality in 2D systems corresponding to~(\ref{eqn:SANN3Dev}) in a 3D system, and so it is necessary to solve the nonlinear equation~(\ref{eqn:SANN2D}) numerically to evaluate the radius $R_i^{(m)}$.
A numerical calculation can be performed using the simple bisection or Newton--Raphson method, which incurs additional computational costs.
An explicit formula to obtain $R_i^{(m)}$ in  2D-SANN and a couple of examples are given in Ref.~\onlinecite{isobe_2016a}. 
The 2D-SANN algorithm proceeds as follows:
\begin{enumerate}
\item Estimate the distances $r_{i,j}$ to all potential neighbors $\{j\}$ from a tagged particle $i$.
\item Sort potential candidates from the neighbors $\{j\}$ by their distance $r_{i,j}$ in increasing order.
\item Start with $m = 3$ (i.e., the minimum number of NNs).
\item Compute $2\pi=\sum_{j=1}^m 2\cos^{-1} ( r_{i,j}/R_ i^{(m)} ) = \sum_{j=1}^m 2\theta_{ij}$, and obtain the $R_i^{(m)}$ numerically by the bisection or Newton--Raphson method.
\item If $R_i^{(m)} > r_{i,m+1}$, then, increment $m$ by $1$ and go back to step 4.
\item Otherwise, $m$ is the number of neighbors for $i$, and $R_i^{(m)}$ is the corresponding radius for the (first) NN.
\end{enumerate}

For the purpose of identifying the index number(s) of NNs, it is only necessary to quantify the sign of the function of $R_i^{(m)}$ defined by the following equation to avoid unnecessary computational costs incurred by evaluating explicit values of $R_i^{(m)}$:
\begin{equation}
f (R_ i^{(m)}) = \sum_{j=1}^m 2\cos^{-1}\! \left( \frac{r_{i,j}}{R_ i^{(m)}} \right) - 2 \pi.
\label{eqn:SANN2Dm}
\end{equation}
The  version of the 2D-SANN algorithm incorporating this function, the 2D-SANNex algorithm, proceeds as follows:
\begin{enumerate}
\item[1$'$--3$'$.] Same as steps 1--3 of  2D-SANN. 
\item[4$'$.] Evaluate $f (R_ i^{(m)})$ by substituting $r_{i,m+1}$ for $m$ into $R_ i^{(m)}$ in short distance order of the potential candidates for neighbors $\{j\}$.
\item[5$'$.] In the case of $f < 0$, increment $m$ by $1$ and go back to step 4$'$.
\item[6$'$.] Otherwise, $m$ is the number of neighbors for $i$.
\end{enumerate}

The computational costs of  2D-SANNex  are much lower than those of  2D-SANN  (see Sec.~\ref{sec:3-2}).
The details of the 2D version of the SANN algorithm, as described above, along with methods to avoid direct solution of the nonlinear equation~(\ref{eqn:SANN2D}), are presented for the first time in this paper.

\begin{figure}[!t] 
\includegraphics[scale=0.08]{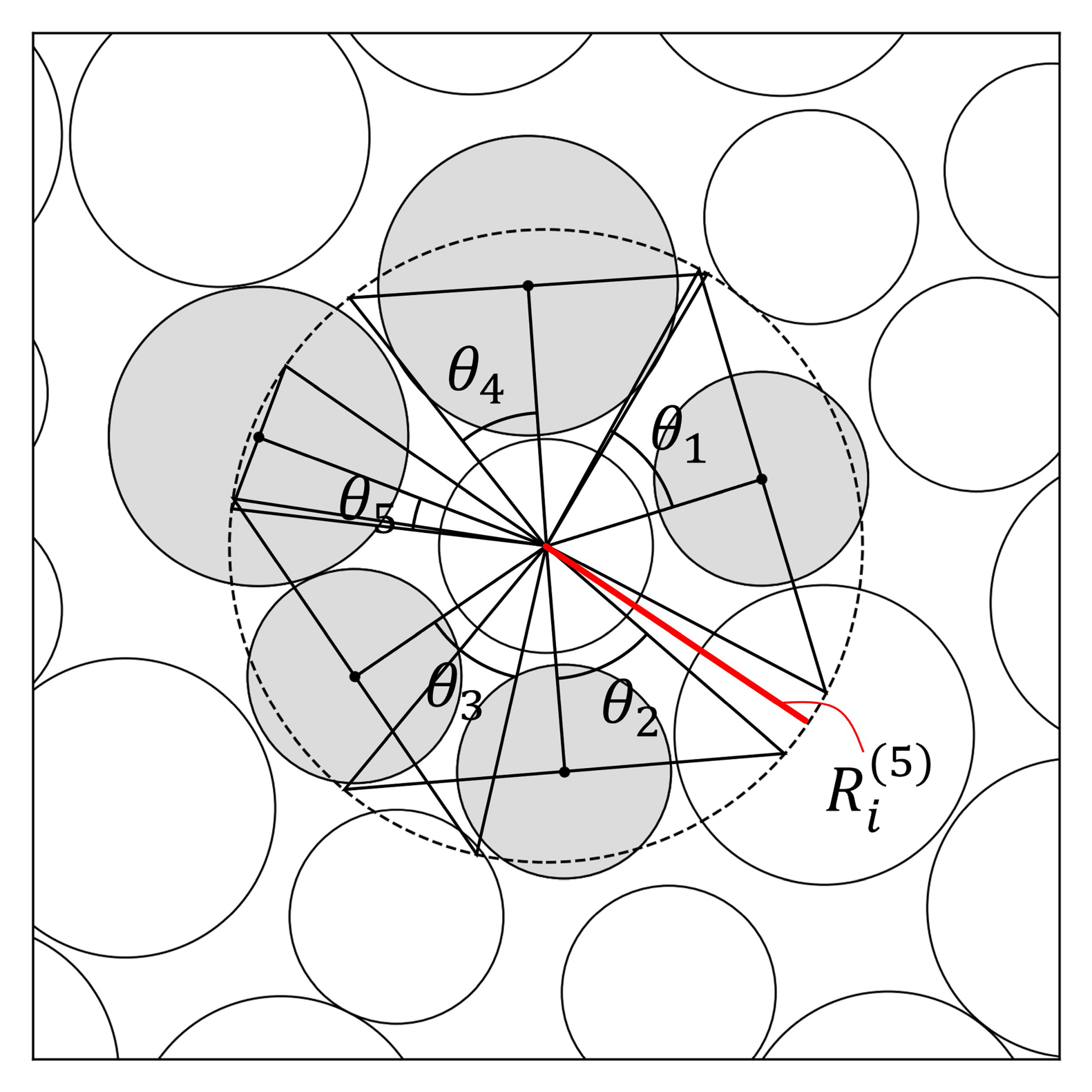}
\caption{Schematic  of 2D-SANN algorithm for a bidisperse system. 
The radius  $R_i^{(m)}$ is increased up to $m=5$, at which Eq.~(\ref{eqn:SANN2D})  holds. 
The potential NNs $\{j\}$ inside the circle of radius $R_i^{(5)}$ from the center of the tagged particle $i$ are recognized as NNs (gray shaded circles).}
\label{fig:SANN_bi}
\end{figure}

Figure~\ref{fig:SANN_bi} shows a typical example of  2D-SANN  as applied to a  bidisperse hard disk system.
Starting from $m=3$ up to $m=5$, the corresponding shell radius $R_i^{(m)}$ is increased such that the right-hand side of Eq.~(\ref{eqn:SANN2D}) exceeds $2\pi$.
The NNs for $i$ can be identified as those lying within the radius $R_i^{(5)}$.

\subsubsection{Higher-order neighbor estimators}
\label{sec:2-1-4}

In the colloid glass experiment reported in Ref.~\onlinecite{yip_2020}, the quasi-voids distributed around second and third neighbors  caused facilitation to induce large particle displacements and initiate  successive hopping motions.
It is therefore crucial to categorize  particles beyond the NN shell in polydisperse systems such as glassy systems. 
However, there are no precise criteria and few algorithms to detect higher-order neighbors.

\begin{figure}[!t] 
\includegraphics[scale=0.13]{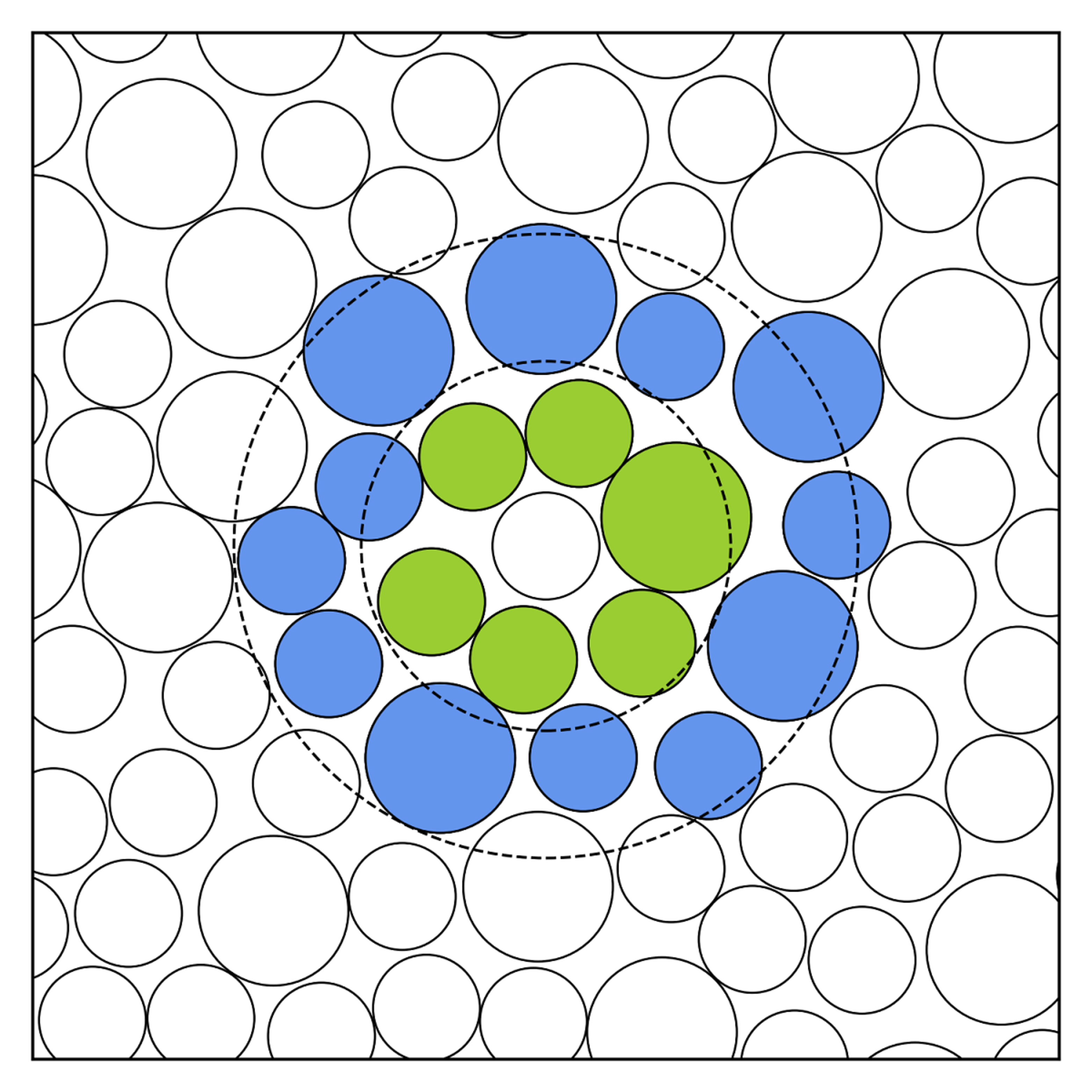}
\caption{Higher-order (up to second) neighbors according to the extended cutoff in a bidisperse system.
The numbers of first (yellow-green) and second (light blue) neighbors are $6$ and $12$, respectively.
}
\label{fig:NN_cutoff_bi_higher}
\end{figure}

The conventional cutoff can easily be extended to a higher-order method using RDFs\cite{isobe_2012}.
To systematically consider  neighbors farther than  nearest neighbors, we define neighbor shells based on the minimum of the RDF for each packing fraction $\nu$, which can be obtained by independent calculations via event-driven molecular dynamics (EDMD)\cite{isobe_1999}.
The second neighbors (next-nearest neighbors) of a tagged particle $i$ can be categorized as the centers of neighbor particles  located at distances from $i$ between $r_{\mathrm{c}}$ and $r^{(2)}_c$, which are estimated by the first and second minima of the RDF.
We can further detect $n$th neighbors of $i$ between $r^{(n-1)}_c$ and $r^{(n)}_c$, which are the cutoff lengths from the $n$th minimum of RDF.
Figure~\ref{fig:NN_cutoff_bi_higher} shows a schematic of the higher-order neighbors around a tagged particle $i$ according to the extended cutoff.
However, in general, the second and further minima of the RDF will be broad, and the presence of some split peaks, even in dense simple liquids, will, by definition, cause profound ambiguity.
Furthermore, it becomes more complicated to consider such minima in bi-/polydisperse and heterogeneous systems. 

Since the SANN algorithm can be applied to heterogeneous systems without parameters, we extend the 2D-SANN algorithm to detect higher-order neighbors.
In the original paper on SANN\cite{meel_2012}, an algorithm for the next (second) neighbors based on SANN is described.
As a first step, the (first) NNs are computed using the original SANN; then, those neighbors are discarded, and SANN is performed again.
However, as shown by the results in Ref.~\onlinecite{meel_2012}, this extension could identify the next-nearest neighbor only partially and did not work well.

Here, we revise the extended SANN algorithm for higher-order neighbors. 
The 2D-SANN equation~(\ref{eqn:SANN2D}) is slightly modified as follows:
\begin{equation}
2 n\pi= \sum_{j=1}^{m} 2 \cos^{-1}\!\left(\frac{r_{i,j}}{R^{(m)}_{i}}\right).
\label{eqn:higherNN_SANN2D}
\end{equation}
where $n$ on the left-hand side  corresponds to the $n$th neighbors, and thus Eq.~(\ref{eqn:higherNN_SANN2D}) reduces to Eq.~(\ref{eqn:SANN2D}) in the case of $n = 1$. 
Before  Eq.~(\ref{eqn:higherNN_SANN2D}) is adopted to identify $n$th neighbors, particles categorized as first to $(n - 1)$th neighbors must be extracted from the candidates for $n$th neighbors, 
otherwise the contribution of the accumulated angles of neighbors located within $(n - 1)$th neighbors would become dominant and  prevent correct detection of $n$th neighbors.
The  extended 2D-SANN algorithm proceeds as follows:
\begin{enumerate}
\item The (first) NNs around a tagged particle $i$ are identified by 2D-SANN, and these are then removed from the list of candidate neighbors.
\item The extended 2D-SANN in the case  $n=2$ described by Eq.~(\ref{eqn:higherNN_SANN2D}) is adopted by increasing the radius $R_i^{(m)}$.
\item The resulting $m$ and $R_i^{(m)}$ are the number of second neighbors and the cutoff radius $R_i^{(m)}$ at which Eq.~(\ref{eqn:higherNN_SANN2D}) holds.
\item The second neighbors are removed, and the same procedure is performed with the extended 2D-SANN in the case  $n=3$.  Further $n$th neighbors can be obtained without ambiguity by repeating these procedures.
\end{enumerate}

Note that  to identify the index number(s) of the $n$th neighbors, it is only necessary to quantify the sign of a function given by a suitably modified Eq.~(\ref{eqn:SANN2Dm}), just as  in 2D-SANNex.

\begin{figure}[!t] 
\includegraphics[scale=0.13]{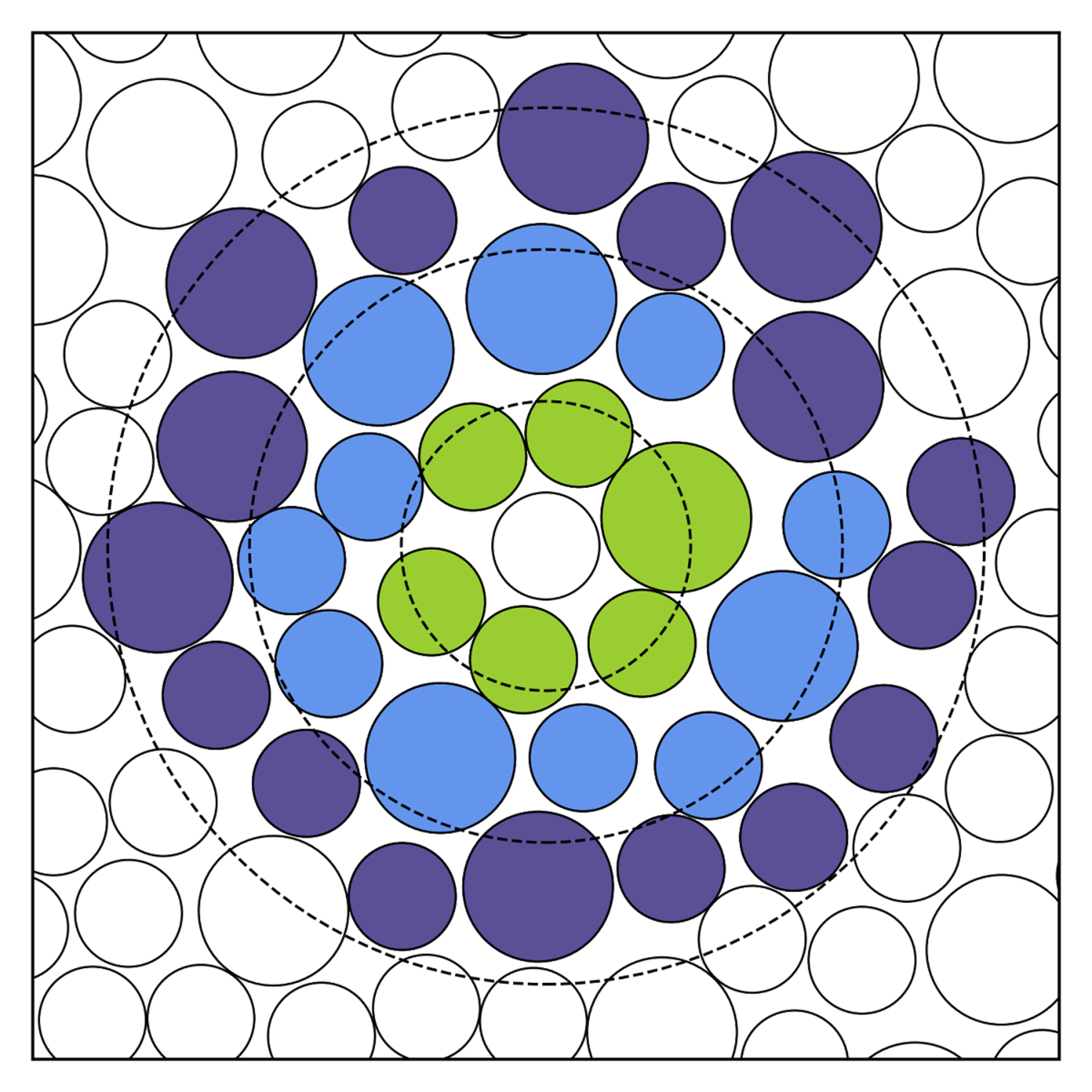}
\caption{Higher-order (up to third) neighbors obtained by extended 2D-SANN for a bidisperse system.
The numbers of first (yellow-green), second (light blue), and third (purple) neighbors are $6$, $11$, and $17$, respectively.}
\label{fig:NN_SANN_bi_higher}
\end{figure}

In bidisperse systems, the results for second neighbors from the extended cutoff and extended 2D-SANN algorithms agree reasonably well, as shown in Figs.~\ref{fig:NN_cutoff_bi_higher} and~\ref{fig:NN_SANN_bi_higher}.
Extended 2D-SANN has the same advantages as 2D-SANN.
Unlike the extended cutoff based on the RDF, the definition of  higher-order neighbors by  extended 2D-SANN is well-defined. 
On the contrary, the cutoff radius of higher-order $r^{(n)}_c$ from the global RDF in the extended cutoff method is a single  parameter for the whole system, and thus this method cannot be used to analyze  local information (e.g., in inhomogeneous systems) precisely.
Higher-order $n$th neighbors can be defined in principle by  extended 2D-SANN even in  polydisperse and  heterogeneous systems just from the information about the local configurations  surrounding each particle.

\subsection{Free volume estimators}
\label{sec:2-2}

\subsubsection{Definitions of free volume and cavity}
\label{sec:2-2-1}

\begin{figure}[!t]
\includegraphics[scale=0.13]{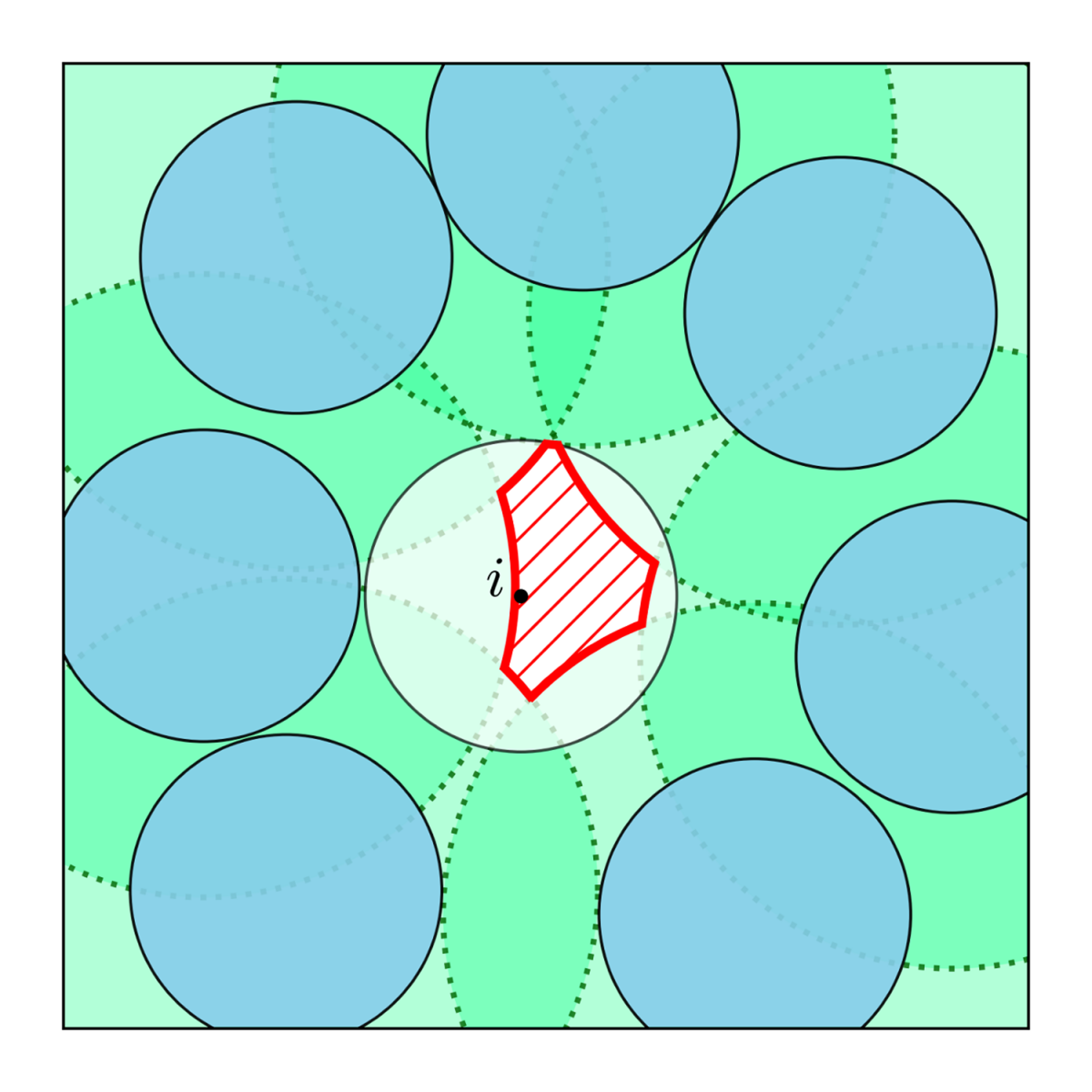}
\caption{Schematic  of the free volume of disk $i$ (shown in  red)}
\label{fig:fv_typical}
\end{figure}

\begin{figure}[!t]
\includegraphics[scale=0.09]{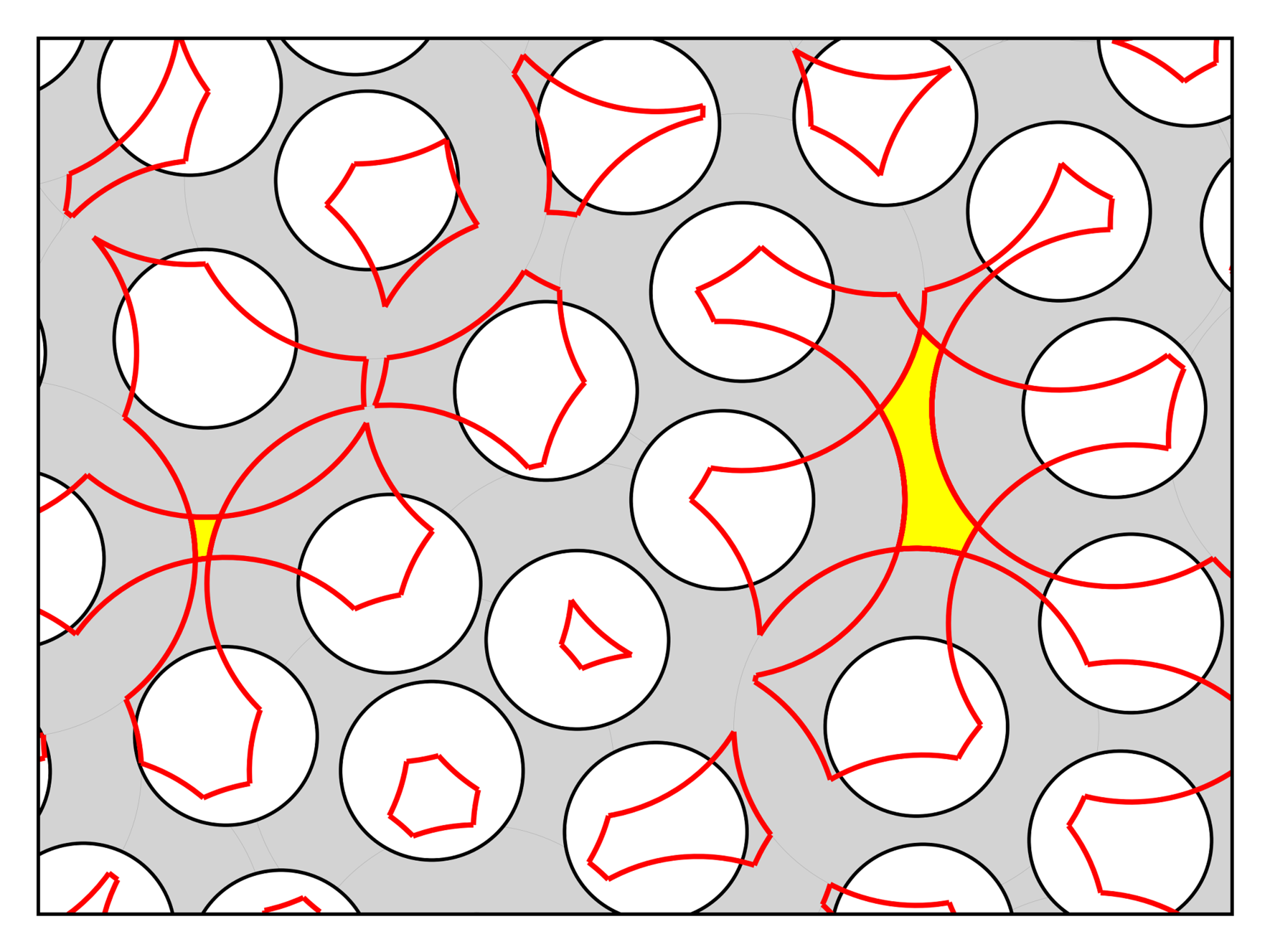}
\caption{Schematic  of  cavities in a hard disk system (shown in yellow).}
\label{fig:cavity_typical}
\end{figure}

The free volume in a 2D system is the region where the center of a tagged disk can move freely if the configuration of the surroundings is fixed, as shown by the red shaded area in Fig.~\ref{fig:fv_typical}.
There are  regions that are inaccessible to the center of the tagged disk $i$, corresponding to the excluded volume due to the presence of the other seven surrounding disks and shown as dotted darker green  circles in Fig.~\ref{fig:fv_typical}.
Note that the free area in a 2D system is also referred to as the ``free volume'' throughout this paper.
In Fig.~\ref{fig:cavity_typical}, the yellow regions are cavities.
The regions bounded by the red curves containing the center of each disk (including the yellow region if adjacent to it) are also described as  free volumes (or free areas) for each disk. The shape of  a free volume $v_{\mathrm{f}}$ can be characterized by the surface area of its bounding planes (i.e., the area of the free surface) $s_{\mathrm{f}}$. Correspondingly, the shape of a free area is characterized by the lengths of its bounding edges. The ratio $s_{\mathrm{f}}/v_{\mathrm{f}}$ is closely related to pressure, as discussed in Sec.~\ref{sec:3-4}.

\subsubsection{Free volume calculation: conventional schemes}
\label{sec:2-2-2}
In  numerical simulations, detecting the set of surrounding particles from which  inaccessible regions are constructed is not trivial, owing  to the complex geometry of the free volume created by the excluded volume. 
Several numerical algorithms for calculating the void space in  dense particle systems or the free volume of a tagged disk have been proposed\cite{hoover_1979, speedy_1991a, rintoul_1995, sastry_1997, sastry_1998}.
The conventional Monte Carlo method with random number generators is simple; however, it is an approximate calculation and has poor convergence with regard to accuracy, especially in highly packed dense systems.
In technical implementations, it is difficult to cover the correct sampling regions, especially in the case of sparse systems or  of anisotropic free volume in inhomogeneous systems. 

Hoover \emph{et al.}\cite{hoover_1979} presented a  rigorous numerical calculation of the free volume in hard disk systems using the circulation to find the intersection point of excluded volume circles. 
Rintoul and Torquato\cite{rintoul_1995} developed a method for void space calculation as an extension of the method of Speedy and Reiss\cite{speedy_1991a}, which is similar to that of Hoover \emph{et al.}\cite{hoover_1979} in the case where  particles can overlap at high density. Void regions are determined by identifying the edges of a disk that are not inside another disk. 
An isolated polygonal area organized by 
the coordinates of intersections between edges of disks
sorted in a clockwise direction is estimated and 
segments of the disks are subtracted from this area.
They also discussed NN distribution functions.
 Voronoi tessellation can also be used for  accurate numerical estimation of the void space and free volume. 
In the late 1990s, Sastry and colleagues\cite{sastry_1997, sastry_1998} implemented an exact calculation method based on  Voronoi polygons and Delaunay triangles in both 2D and 3D systems.
They also generalized their method to the case of polydisperse sphere packing \cite{sastry_1997, maiti_2013, maiti_2014} by radical plane construction\cite{gellatly_1982}, as well as deriving an
 equation of state for mono- and polydisperse hard sphere systems.
However, the computational cost of constructing  Voronoi polyhedra is relatively high, and in 
 the case of polydisperse systems, the computations become complex. 

\subsubsection{Free volume by NELF-A}
\label{sec:2-2-3}
As an alternative algorithm for dense, polydisperse, and larger systems, we implement a simple, efficient, and precise method for categorizing neighbors for enclosing the local free area (called NELF-A), which is easy to implement and can be applied to the dense polydisperse hard disk systems often used as glassy model systems\cite{isobe_2016b}. 
NELF-A focuses on the geometry of intersections of exclusion disks based on a similar strategy to that in Refs.~\onlinecite{hoover_1979, speedy_1991a, rintoul_1995}.
Two adjacent intersections of the excluded volume circles from which  the free volume is constructed (called confirmed intersections) are always angularly adjacent to the set of intersections possessed by each confirmed disk.

The method consists of two stages: the first  identifies the set of confirmed neighbor disks used to construct  the free volume;
the second  calculates the free area using the set of confirmed disks obtained in the first stage.
This method can be applied rigorously not only to monodisperse systems, but to polydisperse systems also.
The first stage is akin to the algorithm presented in Ref.~\onlinecite{hoover_1979}, which employs the intersections of excluded volume circles ordered by angles.
However, NELF-A diverges from that method in more nuanced details, such as the way in which  initial confirmed intersections are identified, rendering NELF-A a more straightforward algorithm. Furthermore, the first stage involves various strategies for determining the appropriate timing to identify confirmed disks (and intersections). The present paper represents the first proposal of such multiple options, and their subsequent benchmarking in polydisperse large-scale systems.

As illustrated in Fig.~2 of Ref.~\onlinecite{rintoul_1995}, the presence of isolated excluded volume circles in empty space results in  failure of the first stage. Consequently, NELF-A is subject to a lower limit on the achievable packing fraction. In the context of the bidisperse system with $64^2$ disks discussed in Sec.~\ref{sec:3-0}, NELF-A proves to be practically viable up to a packing fraction of approximately $\nu \sim 0.450$.

\subsubsection{Detecting confirmed disks for constructing free volume}
\label{sec:2-2-4}
The initial step of NELF-A involves identifying the first intersection that forms the free volume. Subsequently, the particles and intersections that contribute to the free volume are identified in an anticlockwise direction starting from this initial intersection. In this context, the initial intersection can be determined as the nearest intersection. Although there are several algorithms available for such calculations, we introduce three specific algorithms in this paper.

\medskip
\paragraph*{NELF-A by extended 2D-SANNex}
\begin{enumerate}
\item Employ extended 2D-SANNex to detect neighboring disks within the $n$th shells (e.g., $n=2$) for  tagged disk $i$. Include a sufficient number of candidate disks $\{j\}$ to consider the free volume of $i$.
\item Identify the set of all intersections $\mathbf{X}^i_{\{k\}}$ representing excluded volume circles between candidate disks $\{j\}$ for $i$. Calculate these intersections using the positions $\mathbf{r}_i$ and $\mathbf{r}_{\{j\}}$, and their radii $\sigma_i$ and $\sigma_{\{j\}}$.
\item Focus on the closest intersection from $\mathbf{r}_i$, denoted by $\mathbf{X}^i_1$ in $\mathbf{X}^i_{\{k\}}$. $\mathbf{X}^i_1$ is the first confirmed intersection at which the excluded volume circle of the two candidate disks is shared. These two disks become the first and second confirmed disks, labeled as $j_1$ and $j_2$ in angular order from $i$.
\item Focus on the second confirmed disk $j_2$ and intersections $\mathbf{X}^i_1$. The second confirmed intersection $\mathbf{X}^i_2$ is located on the excluded volume circle of $j_2$ in an adjacent anticlockwise order.
\item The $l$th confirmed disk $j_l$ can be identified by the intersection $\mathbf{X}^i_{l-1}$, and the $l$th confirmed intersection $\mathbf{X}^i_l$ can be determined in the same manner as in step 4.
\item Repeat step 5 until the next confirmed disk $j_l$ is equal to $j_1$. 
\end{enumerate}

\begin{figure}[!t]
\includegraphics[scale=0.1]{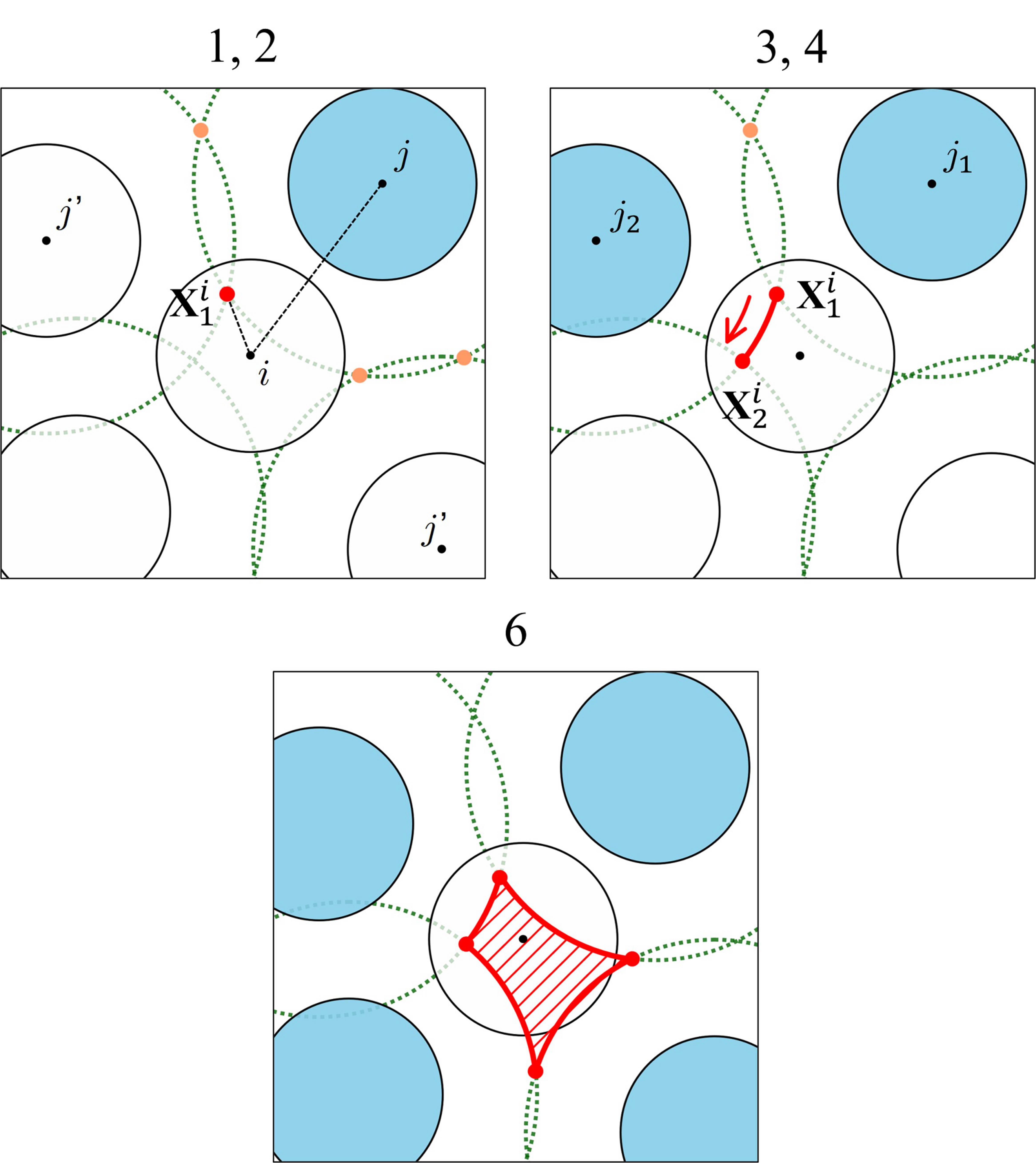}
\caption{Schematic  of NELF-A by cutoff. The particle in the center represents the tagged particle $i$, and the blue particles and red intersections denote confirmed disks and intersections, respectively. The red shaded area represents the free volume of particle $i$.}
\label{fig:fv_fix}
\end{figure}

\paragraph*{NELF-A by cutoff} (see Fig.~\ref{fig:fv_fix})
\begin{enumerate}
\item Find the closest particle $j$ from $\mathbf{r}_i$ and calculate its intersections using the fixed-distance cutoff with a radius $r_{\mathrm{c}}^{\mathrm{N}}=2\sigma_i + \sigma_j + \sigma_{j'}$, where $j'$ is one of the neighbors of $j$. Particles $\{j'\}$ inside such a cutoff radius $r_{\mathrm{c}}^{\mathrm{N}}$ definitely have intersections with $j$.
\item Obtain the closest intersection from $\mathbf{r}_i$ as the first confirmed intersection $\mathbf{X}^i_1$.
\item Determine that the two particles sharing $\mathbf{X}^i_1$ are $j_1$ and $j_2$ in angular order from $i$.
\item Find the intersections of $j_2$ by the fixed-distance cutoff, and define $\mathbf{X}^i_2$ as the intersection next to $\mathbf{X}^i_1$ in an anticlockwise direction.
\item Determine the particle that shares $\mathbf{X}^i_2$ and is not $j_2$ as $j_3$.
\item Repeat steps 4 and 5 until the next confirmed disk is equal to $j_1$. 
\end{enumerate}

\paragraph*{NELF-A by first finding all intersections}
\begin{enumerate}
\item Obtain the intersections $\mathbf{X}^{\{i\}}_{\{k\}}$ for all particles ${\{i\}}$ in the system.
\item Follow the same procedures as in  steps 3--6 of NELF-A by extended 2D-SANNex.
\end{enumerate}

In Fig.~\ref{fig:fv_fix}, each procedure of NELF-A by cutoff is shown.
The efficiency of an algorithm depends on various factors, such as system size, packing fraction, particle dispersions, etc., and the algorithm should be used accordingly.
In Sec.~\ref{sec:3-2}, we discuss  actual calculated efficiencies.

\begin{figure}[!t]
\includegraphics[scale=0.15]{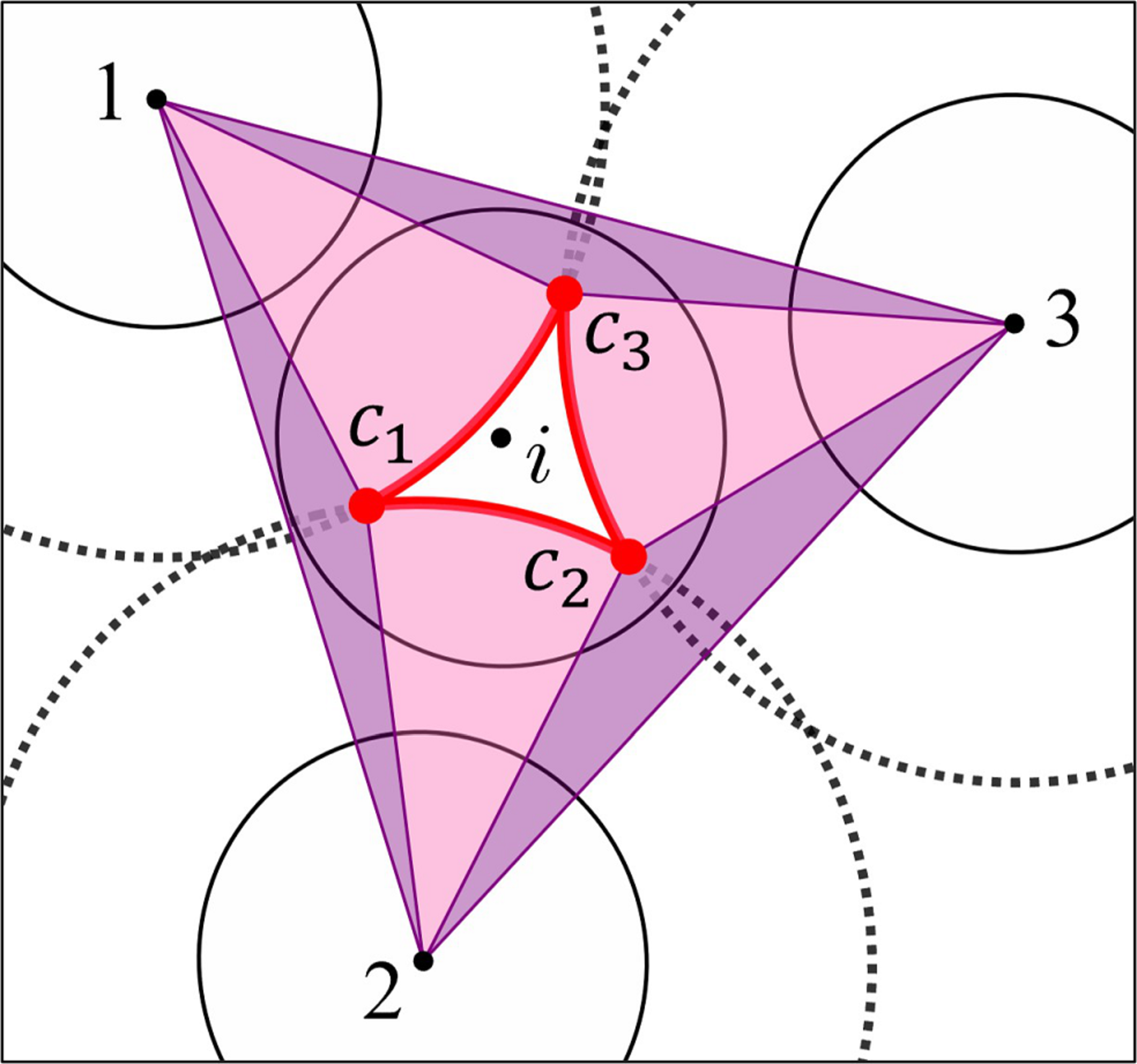}
\caption{The area of the triangle constructed from  the centers of three confirmed disks can be divided into three triangles (purple areas), three sectors (pink areas), and the free volume for disk $i$ (outlined by the red curves).}
\label{fig:fv_calc1}
\end{figure}

\subsubsection{Calculation of free area using confirmed disks}
\label{sec:2-2-5}
For simplicity, we explain how to estimate the free volume constructed by three confirmed disks, labeled $1$, $2$, and $3$. Let $c_1$, $c_2$, and $c_3$ be the intersection points of the exclusion circles of these confirmed disks, as shown in Fig.~\ref{fig:fv_calc1}. 
The free volume of disk $i$ is the area enclosed by the red curves.
Since the triangular area constructed from the centers of the three confirmed disks $1$--$2$--$3$ (labeled as $v_{\mathrm{T}}$) includes the free volume $v^i_{\mathrm{f}}$ for disk $i$, the free volume can be calculated by subtracting the small purple triangles $1$--$2$--$c_1$ ($v_{\mathrm{t}}^1$), $2$--$3$--$c_2$ ($v_{\mathrm{t}}^2$), and $3$--$1$--$c_3$ ($v_{\mathrm{t}}^3$), and the pink sectors $1$-$c_1$-$c_3$ ($v_{\mathrm{s}}^1$), $2$--$c_2$--$c_1$ ($v_{\mathrm{s}}^2$), and $3$--$c_3$--$c_2$ ($v_{\mathrm{s}}^3$); that is, $v^i_{\mathrm{f}}=v_{\mathrm{T}}-\sum_{k=1}^3 (v_{\mathrm{t}}^k+v_{\mathrm{s}}^k)$.
Once the coordinates of the triangle's vertices $\mathbf{X}_{c_k}$ are known, the area can be easily estimated.
The areas of the sectors are also easy to estimate by considering the ratio of the central angle to $2\pi$.

\section{Results and Applications}
\label{sec:3-0}

\subsection{Numerical settings} 
\label{sec:3-1}
We demonstrate the differences between the methods with a typical equilibrated configuration in  systems of both mono- and bidisperse elastic hard disks in two dimensions. In the investigation of  hard sphere/disk models, the development   of computer simulation methods, namely, Monte Carlo (MC)\cite{metropolis_1953} and molecular dynamics (MD)\cite{alder_1959}, represented a milestone, enabling researchers to unveil the mysteries of the phase transitions deriving from the entropic force\cite{wood_1957, alder_1957}(the so-called Alder transition).
In  hard disk systems\cite{alder_1962}, the controversy regarding  the type of phase transitions has now been settled by massive calculations using three modern methods\cite{bernard_2011, engel_2013}.
Hard disk systems are subject to two phase transitions, between fluid, coexisting, hexatic, and solid phases, at around $\nu \sim 0.700$, $0.716$, and $0.720$, respectively.
The free volume of a hard sphere is an essential quantity for understanding phenomena such as  phase transitions and the coarsening process of phase separation (entropic force) in binary mixtures approaching the glass transition\cite{isobe_2016a, isobe_2016b}.

The equilibrated configurations used in the present study were generated by sufficiently long EDMD runs\cite{isobe_1999} at a packing fraction $\nu= N \pi \sigma^2 /A$.
The basic units of the systems are the mass of a single disk $\mu$, the disk diameter $d$ ($= 2\sigma$), and the energy $1/\beta$, from which we derive the unit of time as $d\sqrt{\beta \mu}$, with $\beta=1/(k_{\mathrm{B}}T)$, where $k_{\mathrm{B}}$ the Boltzmann constant  and $T$ is the temperature.

For the monodisperse hard disk system,  disks of radius $\sigma$ were placed in an $L_x \times L_y~(= A)$ rectangular box ($L_y/L_x = \sqrt{3}/2$) with periodic boundary conditions.

For the bidisperse hard disk system (like those often used as simple glassy models\cite{isobe_2016a, isobe_2016b}), $N=(x_0+x_1)N$ additive binary hard disks were considered, with mole fractions  $x_0 = 2/3$ (small) and $x_1 = 1/3$ (large), respectively.
The disks were placed in an $L_x \times L_y~(= A)$ square box ($L_y/L_x = 1$) with periodic boundaries, with the system  prepared for each packing fraction $\nu= N \pi (x_0\sigma_0^2+x_1\sigma_1^2)/A$.
The size ratio was set at $\alpha = \sigma_1/\sigma_0 = 1.4$, where $\sigma_0$ and $\sigma_1$ are the radii of the small and large disks, respectively.

To optimize the codes of the cutoff and 2D-SANN algorithms, we have introduced a grid mapping technique (the exclusive grid particle method)\cite{isobe_1999} for the neighbor list for each particle. 

\begin{table*}[!t]
\caption{Comparison of NN calculations between the fixed-distance cutoff, 2D-SANN, and 2D-SANNex algorithms and of free volume calculations between (a) NELF-A by extended 2D-SANNex, (b) NELF-A by cutoff, and (c) NELF-A by first finding all intersections for mono- and bidisperse systems. }
\label{tab:eff}
\begin{ruledtabular}
\begin{tabular}{lcccccc}
& \multicolumn{6}{c}{CPU time per configuration (ms)} \\ \hline
& \multicolumn{3}{c}{NNs} & \multicolumn{3}{c}{Free volume} \\ \cline{2-4}\cline{5-7}
& Cutoff & 2D-SANN & 2D-SANNex & (a) & (b) & (c) \\ \hline
Monodisperse & $0.332$ & $48.4$ & $4.55$ & $22.0$ & $26.0$ & $9.60$ \\
Bidisperse & $0.823$ & $49.7$ & $9.04$ & $31.8$ & $30.5$ & $2.27\times10^3$
\end{tabular}
\end{ruledtabular}
\end{table*}

\subsection{Efficiency of NN and free volume estimators}
\label{sec:3-2}
To clarify the typical computational costs (or efficiency) of each NN estimator, we measured the run time as a benchmark in a monodisperse hard disk system with parameters $(N, \nu)=(256^2, 0.720)$ and in a bidisperse system with parameters $(N, \nu, x_1, \alpha)=(64^2, 0.720, 1/3, 1.4)$. Both simulations were performed on one core of an Intel Xeon E5-1660, 3.3 GHz.
We compared the elapsed CPU time required to complete the categorization of NNs for one configuration using the fixed-distance cutoff, 2D-SANN, and 2D-SANNex algorithms, as shown in Table~\ref{tab:eff}. We employed the bisection method for 2D-SANN.

In Ref.~\onlinecite{meel_2012}, for simple Lennard--Jones liquid and fcc crystal systems, the Voronoi construction took $24.4$ and $37.7$ times longer to compute, respectively, in comparison with the fixed-distance cutoff. By contrast, the computational cost of SANN was only $1.8$ and $2.4$ times longer, respectively, demonstrating a significant advantage over the Voronoi construction by an order of magnitude.

For our 2D monodisperse systems, both 2D-SANN and 2D-SANNex took $145.8$ and $13.7$ times longer to compute compared with the fixed-distance cutoff, and this trend was the same for bidisperse systems. The longer computation time of 2D-SANN and 2D-SANNex for our systems compared with Ref.~\onlinecite{meel_2012} may be attributed to the presence of nonlinear terms in 2D systems [see Eq.~(\ref{eqn:SANN2D})]. Nonetheless, we consider 2D-SANNex to be more efficient than the Voronoi construction, since the latter took more than $20$ times longer than the fixed-distance cutoff for all systems computed in Ref.~\onlinecite{meel_2012}.

In addition to NN calculations, we also examined the efficiencies of the various NELF-A methods described in Sec.~\ref{sec:2-2-4}, as presented in Table~\ref{tab:eff}. In the case of NELF-A by extended 2D SANNex, neighboring disks were calculated up to the second shells. For monodisperse systems, NELF-A by first finding all intersections was the fastest, completing the calculation in about half the time taken by the other two methods. Conversely, for bidisperse systems, NELF-A by cutoff demonstrated the best efficiency. NELF-A by first finding all intersections was approximately two orders of magnitude slower than the other two methods for bidisperse systems, and it was slower than for monodisperse systems. We attribute this to the strong influence of particle dispersions on NELF-A and the consequent  need to calculate and memorize the intersections of all particles in the system for every dispersion. As discussed in Sec.~\ref{sec:2-2-4}, considering factors such as particle dispersions and the computer's memory capacity, different algorithms are necessary for efficient computing, depending on the situation. In Sec.~\ref{sec:3-4}, NELF-A by first finding all intersections is used for monodisperse systems, while NELF-A by cutoff is used for bidisperse systems.

\subsection{NN estimators}
\label{sec:3-3}
\subsubsection{Local distributions of the number of NNs} 
\label{sec:3-3-1}
\begin{figure*}
\includegraphics[scale=0.09]{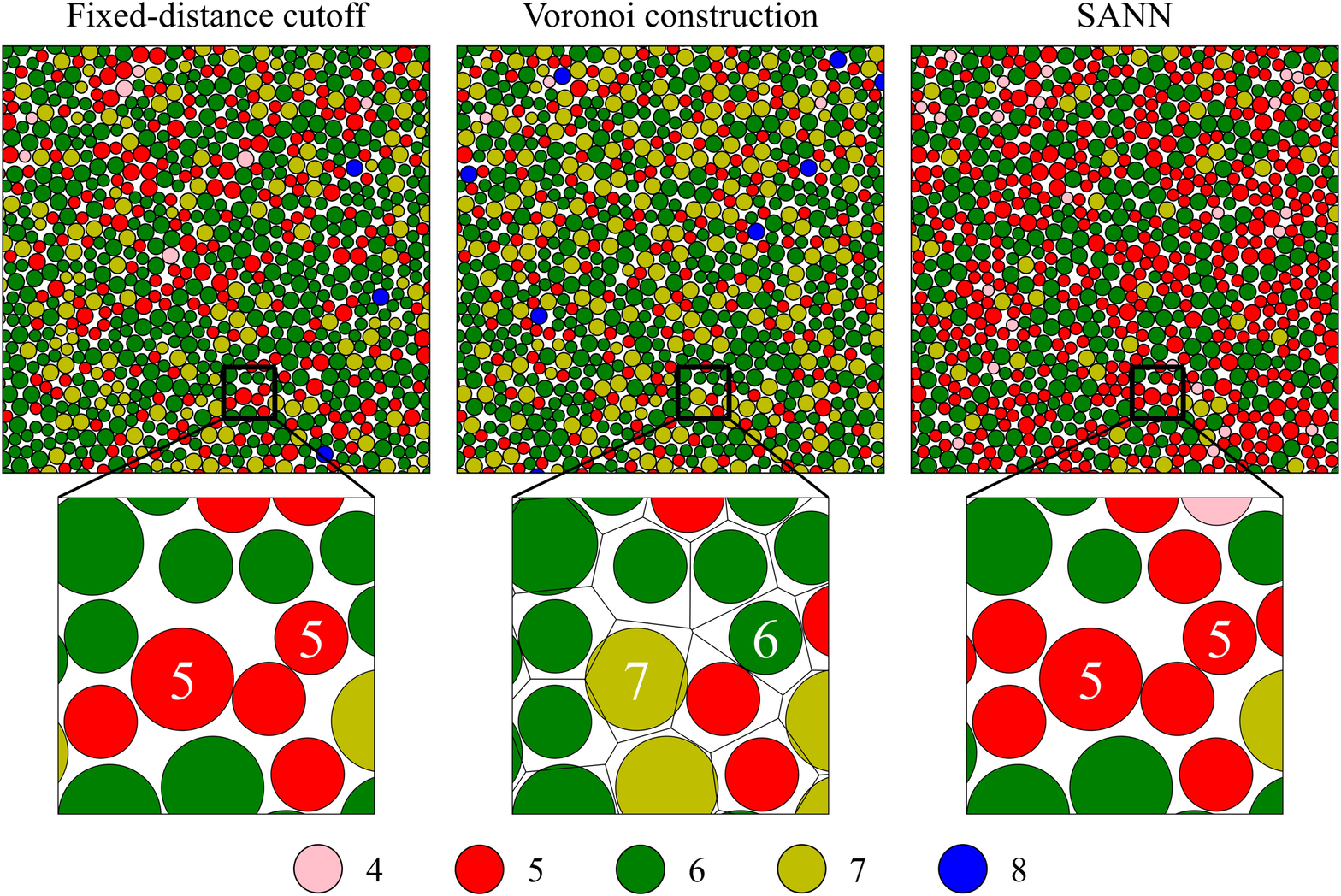}
\caption{Typical configurations of a bidisperse system at $\nu=0.720$ and number of NNs: from left, the fixed-distance cutoff, Voronoi construction, and 2D-SANN.}
\label{fig:NN_cutoff_snap_0720}
\end{figure*}

Snapshots of local structures obtained by NN estimators using a typical configuration of a bidisperse hard disk system reveal the differences between three methods, namely, the fixed-distance cutoff, Voronoi construction, and 2D-SANN algorithms.
Figure~\ref{fig:NN_cutoff_snap_0720} shows  typical configurations of a bidisperse system at $\nu=0.720$ and the number of NNs obtained using  (from left) the fixed-distance cutoff, Voronoi construction, and 2D-SANN.
The disks are filled by colors corresponding to the number of NNs, namely, $4$NNs (pink), $5$NNs (red), $6$NNs (green), $7$NNs (yellow), and $8$NNs (blue).
Note that these visualizations are 
constructed with a width of $31.1\sigma_{\mathrm{e}}$ along both axes of the entire system, and their enlarged views are depicted with a width of $3.7\sigma_{\mathrm{e}}$, where the effective radius $\sigma_{\mathrm{e}}$ is numerically estimated by averaging $\sigma_{ij}$ [$=(\sigma_i+\sigma_j)/2$] for collision partners $(i,j)$ via EDMD\cite{isobe_2016b}.

In the fixed-distance cutoff result, $6$NNs dominate. 
As the packing fraction increases to $\nu=0.760$, we observe an increase in the numbers of $6$NNs and $7$NNs, while the numbers of $4$NNs and $5$NNs decrease, owing to the closer proximity of disks in the denser system.
At $\nu=0.780$, the numbers of $5$NNs and $6$NNs decrease compared with the case of $\nu=0.760$; conversely, the number of $7$NNs increases. For all these packing fractions, 
$6$NNs consistently remains dominant.
With the Voronoi construction, it is observed that the number of $7$NNs is higher compared with the fixed-distance cutoff. Additionally, the number of $8$NNs is also elevated, indicating that the Voronoi construction tends to estimate a greater number of NNs compared with other methods.
When $\nu$ is increased to $0.780$, the number of $6$NNs decreases slightly, while the numbers of $5$NNs and $7$NNs increase.
In 2D-SANN, the majority consist of $5$NNs and $6$NNs at $\nu=0.720$. At $\nu=0.760$, the numbers of $5$NNs and $6$NNs are almost the same as at $\nu=0.720$, and at $\nu=0.780$, the numbers of $5$NNs and $6$NNs remain dominant.

We then focused on the differences in the more localized spatial distributions of NNs between methods. Focusing on the large disk located at the center of each enlarged view in Fig.~\ref{fig:NN_cutoff_snap_0720}, $7$NNs are identified by the Voronoi construction, while the other two methods are identified by $5$NNs. A small part of the edge of a polygon, between the large and small disks (located on the upper right of the large disk), is shared and recognized as an NN, i.e., as topologically connected. However, these disks are located relatively far from the other NNs; i.e., the metric distance is large. The instability in the number of neighbors is expected to be caused by thermal fluctuations, which results in the detection of a relatively large number of NNs. This is identified as one of the disadvantages of the Voronoi construction.

\subsubsection{Probability distribution of the number of NNs}
\begin{figure}[!t]
\includegraphics[scale=0.11]{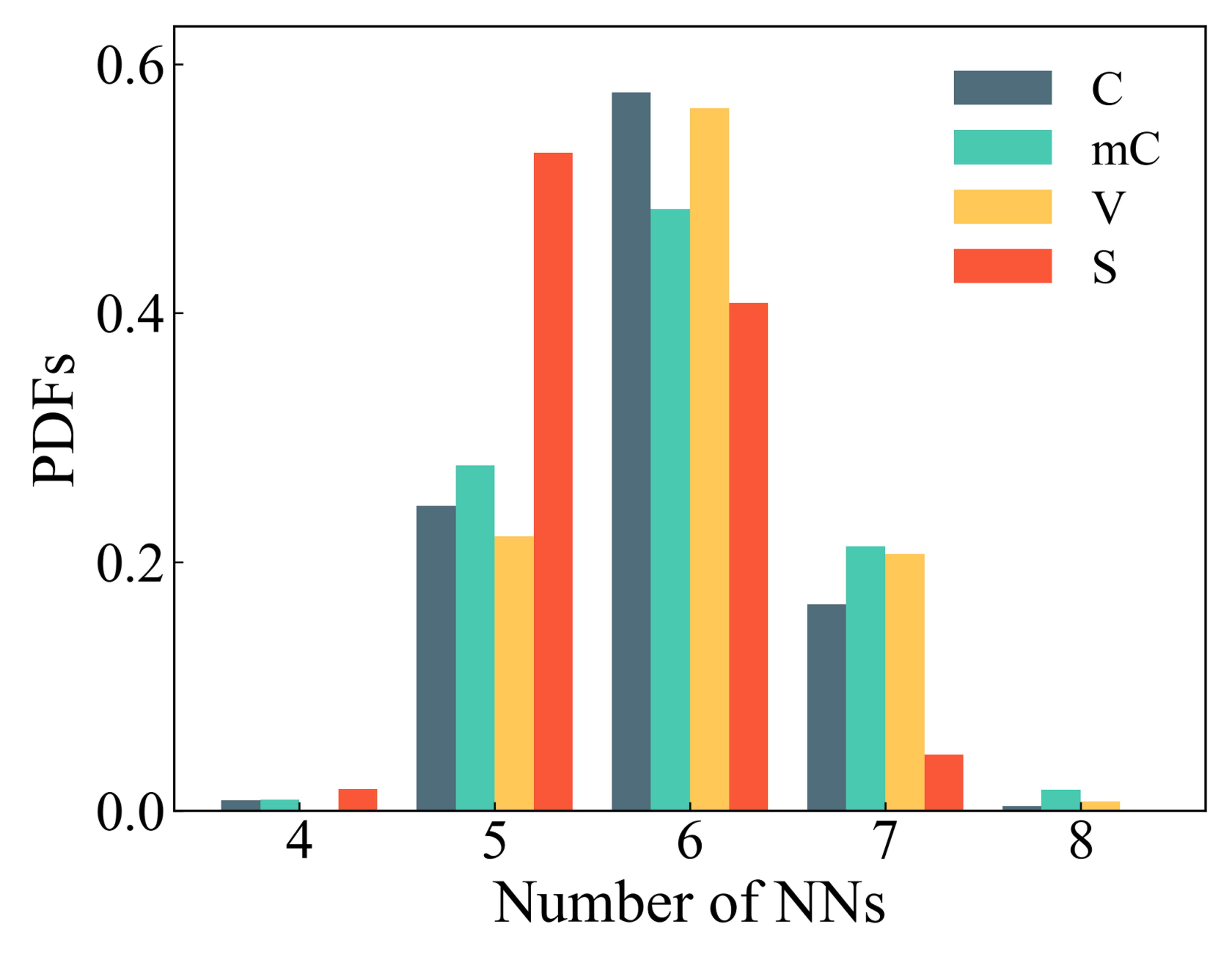}
\caption{PDFs of the number of NNs by the fixed-distance cutoff (C), the modified cutoff (mC), the Voronoi construction (V) and 2D-SANN (S) at $\nu=0.720$ for a bidisperse system.}
\label{fig:NN_PDF_0720}
\end{figure}

The probability distribution functions (PDFs) of the number of NNs in bidisperse hard disk systems at $\nu=0.720$ were calculated using $300$ frame-equilibrated configurations at $(N, \nu)=(64^2, 0.720)$, as shown in Fig.~\ref{fig:NN_PDF_0720}. Here,
the number of NNs is estimated by the fixed-distance cutoff (C), the modified cutoff (mC), Voronoi construction (V), and 2D-SANN (S), respectively. 
A notable difference with 2D-SANN compared with the other methods is the case of $5$NNs has the highest probability, whereas the case of $6$NNs is most probable according to the other methods.
This trend remains consistent at $\nu=0.760$ and $0.780$.
Additionally, with the modified cutoff method, where multiple cutoff distances corresponding to each particle pair are introduced, it is observed that the numbers of $5$NNs and $7$NNs are higher compared with the fixed-distance cutoff.
Furthermore, the Voronoi construction identifies more $7$NNs and fewer $5$NNs than the fixed-distance cutoff.
This observation that a relatively large number of NNs are detected is expected, since  disks located at large metric distances are topologically shared with the short edges.

\begin{figure}[!t]
\includegraphics[scale=0.11]{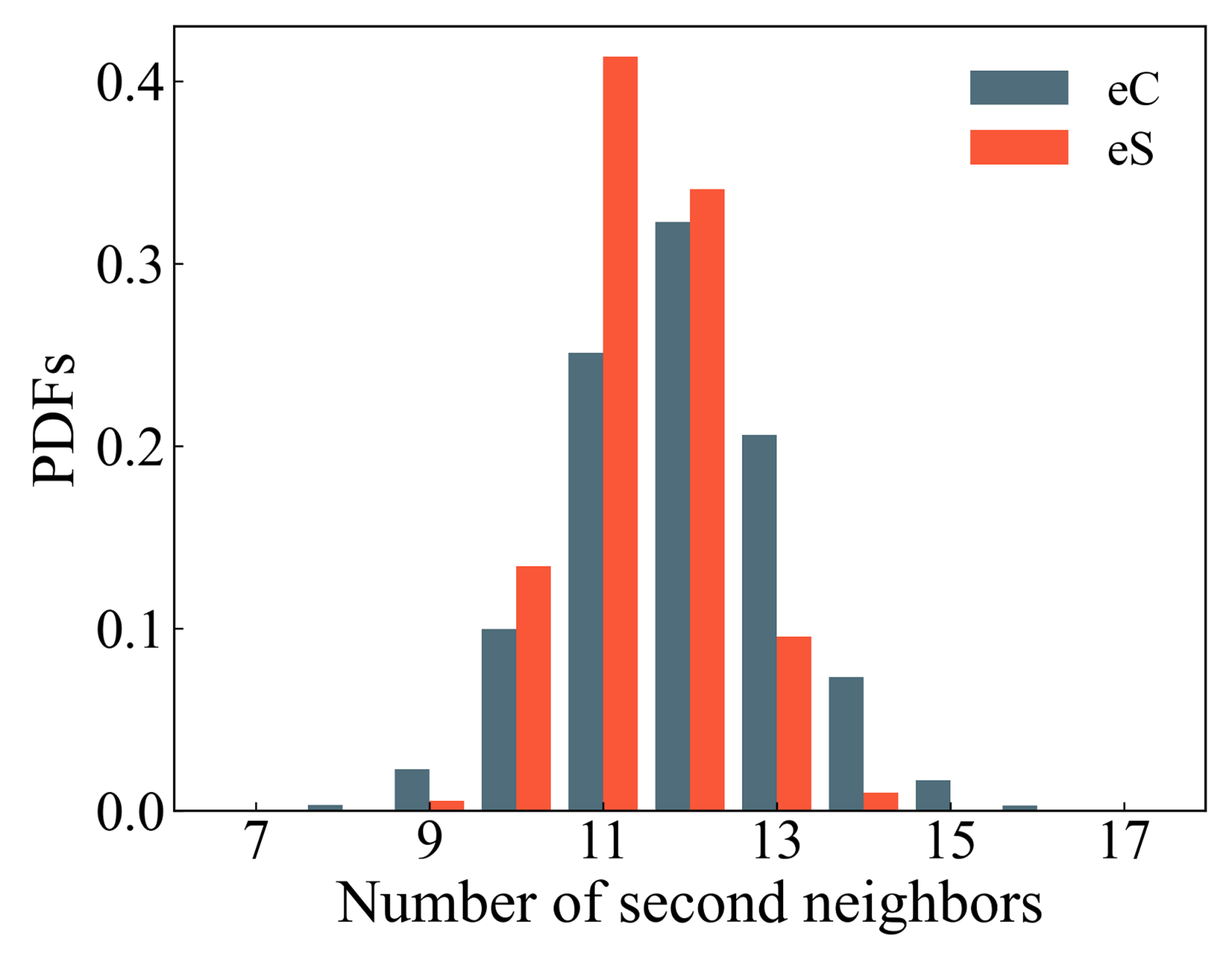}
\caption{PDFs of the number of second neighbors according to the extended cutoff method (eC) and extended 2D-SANN (eS)  at $\nu=0.720$ for a bidisperse system.}
\label{fig:NN_higher_PDF_0720_CS}
\end{figure}

Figure~\ref{fig:NN_higher_PDF_0720_CS} shows a comparison of the PDFs of the number of second neighbors according to the extended cutoff method\cite{isobe_2012} (eC) and extended 2D-SANN (eS) 
(see Sec.~\ref{sec:2-1-4} and Figs.~\ref{fig:NN_cutoff_bi_higher}
and~\ref{fig:NN_SANN_bi_higher}).
With the extended cutoff, from Fig.~\ref{fig:rdf-bi}, the cutoff radii for the first and second shells are set as $r_{\mathrm{c}}^{\ast}=r_{\mathrm{c}}/(2\sigma_{\mathrm{e}})=1.49$ and $2.52$, respectively. 
Extended 2D-SANN is found to give a smaller value for the most frequent number of second neighbors compared with the extended cutoff, and this trend is similar to the results for the (first) NNs.
Additionally, as the particle packing fraction increases  to $\nu=0.780$, in the extended cutoff method, $12$NNs becomes more dominant,  whereas, with extended 2D-SANN, the probabilities of $11$NNs and $12$NNs become almost the same, indicating that both $11$NNs and $12$NNs become dominant as the packing fraction increases.

\subsubsection{Discussion of NN estimators}
\label{sec:3-3-3}
In experiments such as those on colloidal and granular systems, it is impossible to prepare a perfect monodisperse system, owing to the small fluctuations in the size of colloids.
Bidisperse systems are often used to model  glassy systems to avoid crystallization.
In those systems, the  conventional methods, i.e., cutoff and Voronoi construction, have difficulties in identifying NNs, summarized below.

With the fixed-distance cutoff, a suitable metric cutoff distance $r_{\mathrm{c}}$ must be independently estimated for each system with a different packing fraction, following a long calculation for generating the RDF and determining the precise location of its first minimum.
However, in a bidisperse system, the first peak of the RDF is split into three.
Furthermore, in a polydisperse system, not only will the first peak of the RDF  be split, but the second peak will be diffuse even in the case of a dense liquid.
The use of a modified cutoff by independently considering RDFs with different kinds of pairs would be more suitable for bidisperse systems, but this requires three cutoff radii.
The ambiguity in the positions of peaks when determining the cutoff length imposes substantial limitations on this method in the case of polydisperse/inhomogeneous systems.

In bidisperse systems, since the pairs of large and small disks do not have the same radii, the line of the perpendicular bisector between them in the Voronoi construction often intersects with the edge of a circle of the larger disk (see Figs.~\ref{fig:voronoi_mono_bi} and~\ref{fig:NN_cutoff_snap_0720}).
This causes  errors in the construction of polygonal cells for a large disk.
Therefore, a radical plane construction\cite{gellatly_1982} for  polydisperse systems has been proposed and applied\cite{sastry_1997, maiti_2013, maiti_2014}. 
Another difficulty is that if two Voronoi polygons share a small side, a pair of disks will be detected as NNs even when they are separated by a large  metric distance, and this 
can be the cause of the instability under thermal fluctuations mentioned in Ref.~\onlinecite{meel_2012}.
Furthermore, computational costs are relatively high, and a special numerical treatment is required to construct space-filled tessellations in the case of an open boundary.

In SANN, each disk has its individual (local) cutoff distance, which is determined by the topological geometry of the configuration of neighbors around a tagged disk. In the 2D version, the sum of all solid angles associated with the neighbors adds up to $2\pi$.
The SANN algorithm is both parameter-free and scale-free and is  robustness to thermal fluctuations.
These features make it suitable for application to  polydisperse or inhomogeneous systems without the need for any adjustable parameters.
Individual SANN cutoff lengths can be used to define the local density (void structures or softness).
The computational cost of 2D-SANNex is found to be low, which allows on-the-fly simulations.

\subsection{Pressure estimators}
\label{sec:3-4}
One practical application of the free volume is in pressure calculations.
In this subsection, we validate  NELF-A by comparing pressure values obtained through the free volume approach with thhose obtained using an alternative method in a hard disk system.
By showing the merit of  pressure calculations based on the free volume, we aim to confirm the practicality of NELF-A, including its efficiency.

The conventional pressure calculation method in a hard disk system is based on a dynamical expression derived via the virial theorem\cite{alder_1958, erpenbeck_1977, alder_1960, hoover_1967, engel_2013}. The dimensionless pressure $P^\ast$ in two dimensions is given by
\begin{equation}
P^\ast=\beta P(2\sigma)^2=\frac{4\nu}{\pi}\!\left(1-\frac{\beta \mu}{2Nt}\sum_{N_{\mathrm{c}}} b_{ij}\right),
\label{eqn:Pb}
\end{equation}
where $t$ is time in the  units of the simulation and $N_{\mathrm{c}}$ is the collision number.
The collision force $b_{ij}=\mathbf{r}_{ij}\cdot\mathbf{v}_{ij}$ is given by the scalar product between the relative positions $\mathbf{r}_{ij}$ and the relative velocities $\mathbf{v}_{ij}$ of the collision partners $(i,j)$.
In a monodisperse hard disk system at equilibrium, Eq.~(\ref{eqn:Pb}) can be rewritten as
\begin{equation}
P^\ast=\frac{4\nu}{\pi}\!\left(1+\frac{\sigma\sqrt{\pi\beta \mu}}{2}\Lambda\right),
\label{eqn:PL}
\end{equation}
where $\Lambda$ is the collision rate.\cite{erpenbeck_1977, hoover_1967, engel_2013}

Alternatively, pressure can  be expressed in terms of the ratio of free surface area to free volume for each tagged particle $i$, $s_{\mathrm{f}}^i/v_{\mathrm{f}}^i$. Such  expressions have been proposed and applied to  various equilibrated monodisperse hard sphere systems. \cite{maiti_2013, maiti_2014, schindler_2015, hoover_1972, speedy_1980, speedy_1981, speedy_1988, speedy_1991b}
Hoover \emph{et al.}\cite{hoover_1972} derived the following expression for the pressure under the assumption that $\Lambda$ in Eq.~(\ref{eqn:PL}) must be proportional to the ratio $s_{\mathrm{f}}^i/v_{\mathrm{f}}^i$:
\begin{align}
P^\ast&=\frac{4\nu}{\pi}\!\left(1+\frac{\sigma}{2N}\sum_{i=1}^N\left\langle\frac{s^i_{\mathrm{f}}}{v^i_{\mathrm{f}}}\right\rangle\right)\nonumber\\
&=\frac{4\nu}{\pi}\!\left(1+\frac{\sigma}{2N_{\mathrm{s}}}\sum_{i=1}^{N_{\mathrm{s}}}\frac{s^i_{\mathrm{f}}}{v^i_{\mathrm{f}}}\right),
\label{eqn:PFV_mono}
\end{align}
where $\langle s_{\mathrm{f}}^i/v_{\mathrm{f}}^i\rangle=(1/\Omega)\sum_{\Omega} s_{\mathrm{f}}^i/v_{\mathrm{f}}^i$ is the ensemble average of the ratio, with $\Omega$ being the configuration number  and $N_{\mathrm{s}}(=\Omega\times N)$  the total number of sample particles.
Subsequently, Speedy\cite{speedy_1980, speedy_1981} independently derived an expression for the pressure expression that was the same as Eq.~(\ref{eqn:PFV_mono})  by noting that the pair distribution function is equal to the configuration average of the ratio, $\langle s_{\mathrm{f}}^i/v_{\mathrm{f}}^i\rangle$.
Speedy\cite{speedy_1988, speedy_1991b} also rederived this expression on the basis of the statistical properties of cavities in the thermodynamic limit.
In  polydisperse systems, the radius $\sigma$ in the pressure expression~(\ref{eqn:PFV_mono}) can be simply replaced by the radius of the independent particle $\sigma_i$.
Based on Speedy's argument, Corti and Bowles\cite{corti_1999} presented an  explicit  expression for the pressure in the polydisperse case:
\begin{equation}
P^\ast=\frac{4\nu}{\pi}\!\left(1+\frac{1}{2N_{\mathrm{s}}}\sum_{i=1}^{N_{\mathrm{s}}}\sigma_i\frac{s^i_{\mathrm{f}}}{v^i_{\mathrm{f}}}\right)
=\frac{1}{N_{\mathrm{s}}}\sum_{i=1}^{N_{\mathrm{s}}} p_i^\ast,
\label{eqn:PFV}
\end{equation}
where $p_i^\ast$ is the dimensionless partial  pressure:
\begin{equation}
p_i^\ast=\frac{4\nu}{\pi}\!\left(1+\frac{1}{2}\frac{s_{\mathrm{f}}^i}{v_{\mathrm{f}}^i}\sigma_i\right).
\label{eqn:Pp}
\end{equation}
Note that
the partial pressure expression given by Eq.~(\ref{eqn:Pp}) is defined solely  on the basis of the local environment of a single particle at a specific time.
This characteristic renders this expression valuable for application to nonequilibrium systems, such as relaxation processes and time-dependent phenomena, as well as to inhomogeneous systems.
On the one hand, in dynamic pressure expressions, it is necessary to accumulate collision forces over the duration in an equilibrium system.
On the other hand, a static pressure expression using the pair distribution function requires estimation based on the positions of all particles within the system.
The ability to estimate pressure is limited to homogeneous equilibrium systems that do not depend on time in either of these cases.
Recently, Eq.~(\ref{eqn:PFV}) has been applied to protein structure analysis\cite{maiti_2013}, jammed hard sphere packings\cite{maiti_2014}, and specific treatments of  data sets of colloidal diameter that are incomplete owing to experimental limitations\cite{schindler_2015}.

\subsubsection{Efficiency and precision of pressure estimators}

In this subsection, we compare the two pressure expressions given by Eqs.~(\ref{eqn:Pb}) and~(\ref{eqn:PFV})  from the practical aspect of numerical simulation by the EDMD and NELF-A methods, respectively. We label the method based on Eq.~(\ref{eqn:Pb}) and using EDMD as $\mathcal{A}$ and the method based on Eq.~(\ref{eqn:PFV}) and using NELF-A as $\mathcal{B}$.
Note that in $\mathcal{B}$, configuration samples are prepared for collection from the   EDMD trajectories  every 100 mean free times.

\begin{figure*}
\subfloat[]{\includegraphics[scale=0.15]{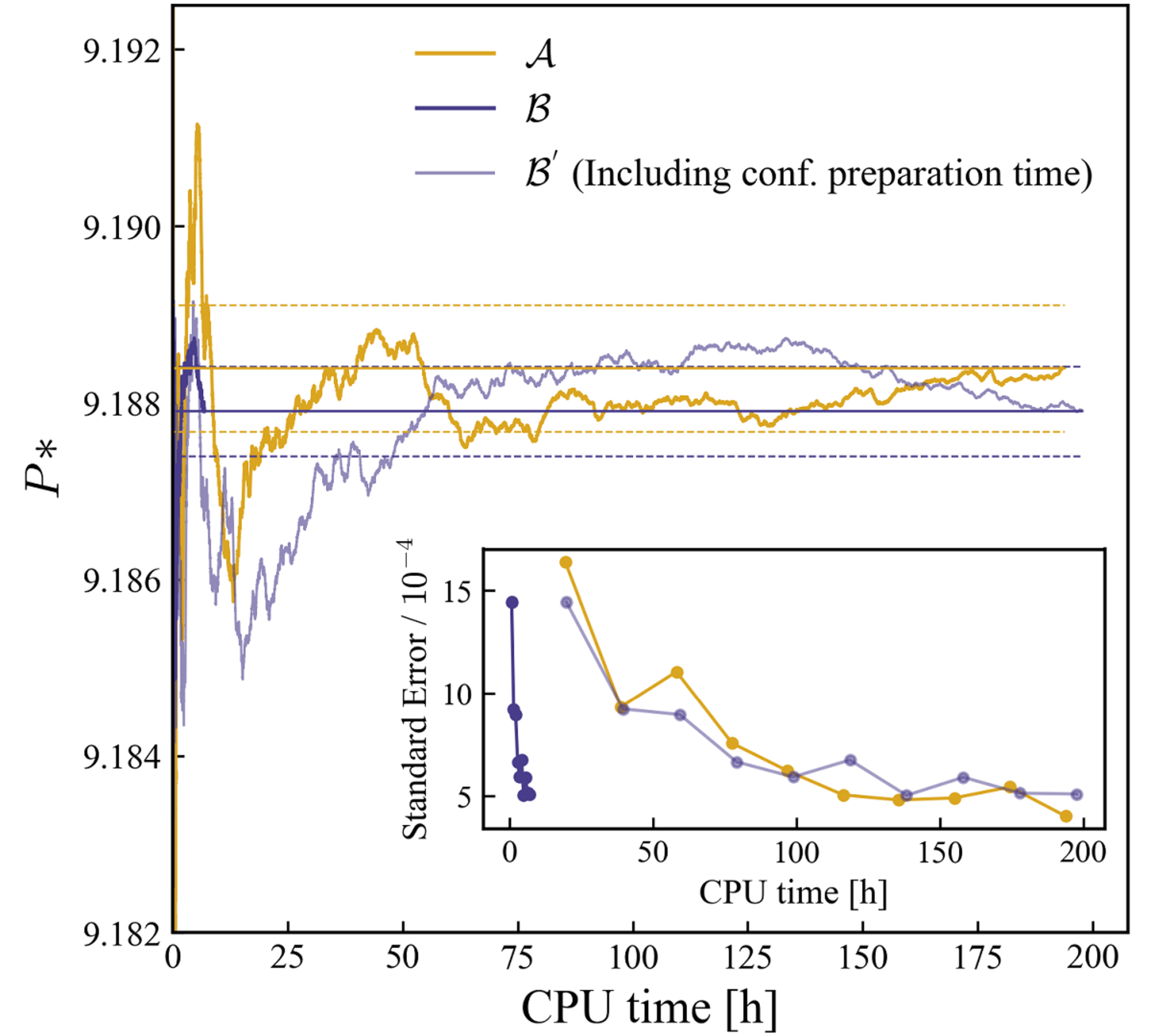}}
\subfloat[]{\includegraphics[scale=0.15]{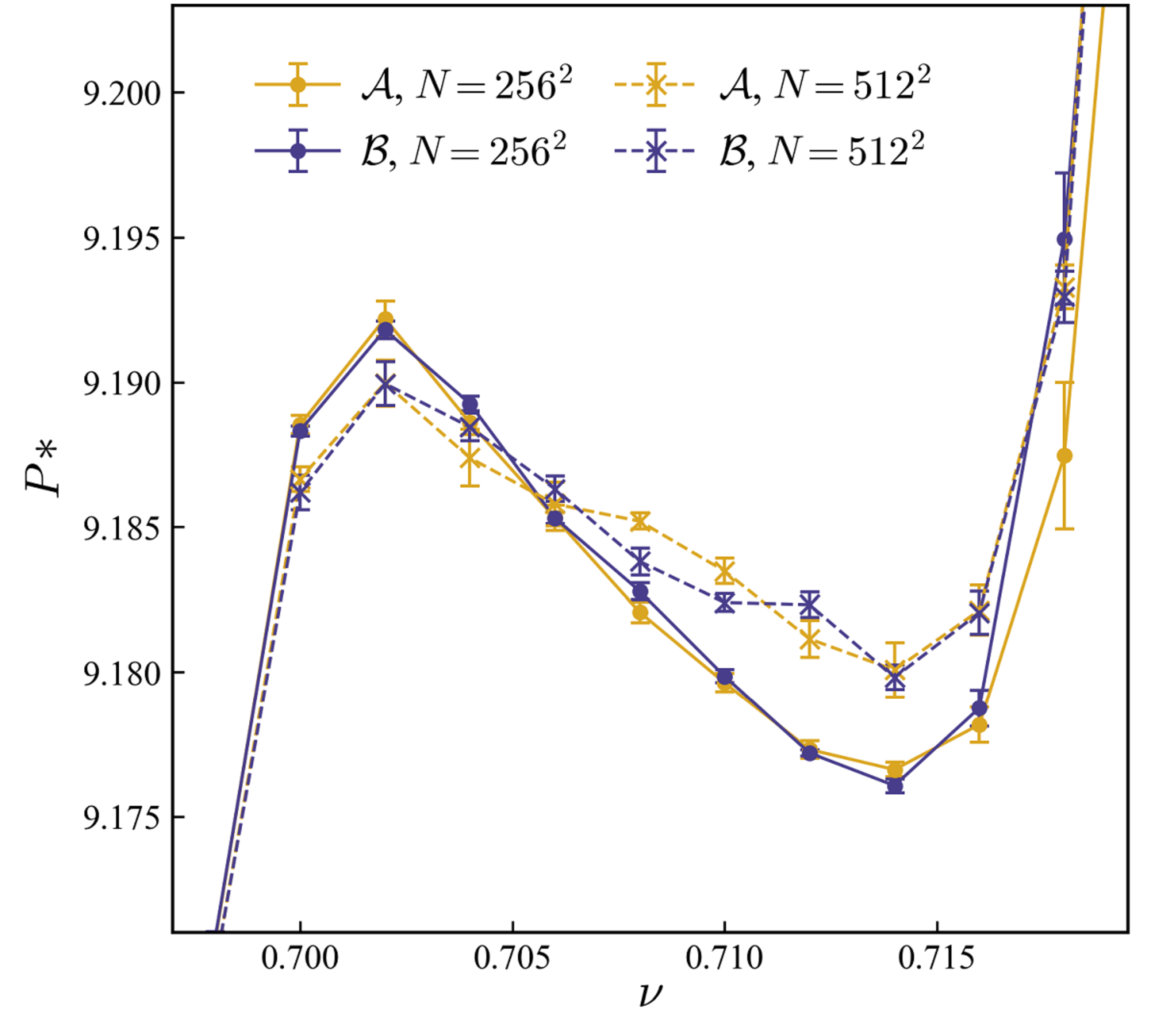}}
\caption{(a) Comparison of evolving pressure  between methods  $\mathcal{A}$, $\mathcal{B}$, and $\mathcal{B}'$  in terms of CPU time for a monodisperse hard disk system at $(N,\nu)=(256^2,0.700)$.
Calculations were performed up to $N_{\mathrm{c}}/N=8.0\times10^6$ for $\mathcal{A}$ and $N_{\mathrm{s}}/N(=\Omega)=1.6\times10^5$ for $\mathcal{B}$.
The inset shows the time evolution of the corresponding standard errors.
(b) Equation of state of a monodisperse system according to methods $\mathcal{A}$ and $\mathcal{B}$ for $N=256^2$ and $512^2$, where  close agreement of the pressures can be observed.
}
\label{fig:P_evol}
\end{figure*}

The accuracy of pressure estimation is improved by  accumulation up to a sampling number $N_{\mathrm{c}}$ in Eq.~(\ref{eqn:Pb}) and $N_{\mathrm{s}}$ in Eq.~(\ref{eqn:PFV}).
Figure~\ref{fig:P_evol}(a)  compares the evolving pressure and the corresponding standard error  with  sampling number for methods   $\mathcal{A}$ and $\mathcal{B}$ in  terms of CPU time for a monodisperse hard disk system with $(N, \nu)=(256^2, 0.700)$.
All simulations were performed using an Intel Xeno E5-1660 with a clock speed of 3.3 GHz.
The  evolving pressures obtained by methods $\mathcal{A}$ and $\mathcal{B}$, respectively, converge to the same value within an error bar after 200 CPU hours.
Method $\mathcal{B}$ appears to be much more efficient than  $\mathcal{A}$; however, it requires independent configurations in the equilibrium, which must be prepared by  EDMD or some other efficient Monte Carlo method\cite{krauth_2021}.
To provide a fair comparison, we also considered the additional computational cost for the preparation of independent configurations and show the total cost as $\mathcal{B}'$ in Fig.~\ref{fig:P_evol}(a).
We found that the computational bottleneck in obtaining accurate pressures for these dense systems lies in sampling the independent configurations, rather than in the pressure calculation itself by NELF-A.

A comparison of the pressures for each packing fraction $\nu$ (i.e., the equation of state) for a monodisperse system obtained by  methods $\mathcal{A}$ and $\mathcal{B}$ is  shown in Fig.~\ref{fig:P_evol}(b).
The pressure from method $\mathcal{A}$ here is depicted using the data from Fig.~3 of Ref.~\onlinecite{isobe_2016a}.
In the case of method $\mathcal{B}$, the pressure was calculated for two parameter sets: $(N, N_{\mathrm{s}}/N, N_{\mathrm{conf}}) = (256^2, 10^5, 5)$ and $(512^2, 10^4, 10)$, where $N_{\mathrm{conf}}$ is the number of independent initial configurations used in executing the EDMD simulation.
We observe that the pressures from methods $\mathcal{A}$ and $\mathcal{B}$ show  reasonably good agreement within an error bar, confirming the validity of NELF-A.
Furthermore, even in a bidisperse system, the pressures from both methods almost coincide for various packing fractions.

We have also compared the efficiency of the two pressure calculation methods for different system sizes and particle dispersions by considering  the accumulated sampling numbers $N_{\mathrm{c}}$ and $N_{\mathrm{s}}$, as shown in Table~\ref{tab:Peff}.
Simulations were again performed for a bidisperse system [$(\nu, x_1, \alpha)=(0.700, 1/3, 1.4)$] as a representative of polydisperse systems.
In all situations, the calculation efficiency with method $\mathcal{A}$ is better than that with $\mathcal{B}$.
It is important to note, however, that this result is only for the efficiency of accumulating $N_{\mathrm{c}}$ and $N_{\mathrm{s}}$; therefore, when the factor of convergence speed as shown in Fig.~\ref{fig:P_evol}(a) is taken into account, it cannot simply be claimed that  calculation with method $\mathcal{B}$ is inefficient.

\begin{table}[!t]
\caption{Comparison of pressure calculations between methods $\mathcal{A}$ and $\mathcal{B}$ for different system sizes and particle dispersions. The accumulated sampling numbers $N_{\mathrm{c}}$ and $N_{\mathrm{s}}$ per millisecond of CPU time are shown.}
\label{tab:Peff}
\begin{ruledtabular}
\begin{tabular}{lcccc}
& \multicolumn{4}{c}{$N_{\mathrm{c}}$ or $N_{\mathrm{s}}$/CPU time (ms$^{-1}$)} \\ \cline{2-5}
& \multicolumn{2}{c}{Monodisperse} & \multicolumn{2}{c}{Bidisperse} \\ \cline{2-3}\cline{4-5}
Method& $N=256^2$ & $N=512^2$ & $N=64^2$ & $N=128^2$ \\ \hline
 $\mathcal{A}$ & $751$ & $460$ & $782$ & $701$ \\
 $\mathcal{B}$ & $275$ & $223$ & $139$ & $136$ 
\end{tabular}
\end{ruledtabular}
\end{table}

\section{Concluding Remarks}
\label{sec:4-0}
In this paper, we have discussed  different the methods for nearest neighbor and free volume estimation, including the higher-order treatments of  polydisperse hard disk systems.
We have clarified the difficulties that arise with  conventional methods using cutoffs and Voronoi constructions in local microscopic environment analyses of molecular configurations.
We have confirmed that the  solid-angle-based nearest-neighbor (SANN) algorithm is able to overcome the difficulties encountered by conventional methods, even for polydisperse hard disk systems.
We have also presented  an explicit definition for evaluating higher-order neighbors, and have proposed a simple, efficient, and precise method for categorizing neighbors for enclosing the local free area (NELF-A) in the case of dense polydisperse hard disk systems.
The  approaches proposed here for the determination of nearest neighbors and the free volume  provide crucial information to elucidate the dynamics of slow relaxation and hopping motion, especially in  densely packed systems.

In the future, we aim to further explore the applications of NELF-A by focusing on two main objectives: (i) developing a methodology for deriving inherent structures in hard disk systems using NELF-A, and (ii) extending the applicability of NELF-A to systems with various boundary conditions and three or more dimensions.

\begin{acknowledgments}
The authors are grateful to Professors W. G. Hoover, C.-H. Lam, and C.-T. Yip for stimulating and fruitful discussions on this topic. Special thanks are extended to Ms. Nur Farahana and Hiroko Ichikawa for providing preliminary results for this project.
M.I. was supported by JSPS KAKENHI Grant Nos.~20K03785 and 23K03246. Part of the computations were performed using the facilities of the Supercomputer Center, ISSP, University of Tokyo.
This research was conducted within the context of the International Research Project ``Non-Reversible Markov Chains, Implementations and Applications.'' 
\end{acknowledgments}

\section*{AUTHOR DECLARATIONS}

\subsection*{Conflict of Interest}
The authors have no conflicts to disclose.

\subsection*{Author Contributions}
{\bf Daigo Mugita}: Conceptualization (equal); Data Curation (lead); Formal Analysis (lead); Investigation (lead); Methodology (lead); Project Administration (equal); Software (lead);  Validation (lead); Visualization (lead); Writing - Original Draft (lead); Writing - Review and Editing (lead).
{\bf Kazuyoshi Souno}: Data Curation (equal); Formal Analysis (lead); Investigation (equal); Methodology (lead); Software (lead);  Validation (equal); Visualization (equal); Writing - Original Draft (equal).
{\bf Hiroaki Koyama}: Data Curation (equal); Formal Analysis (lead); Investigation (equal); Methodology (lead); Software (lead);  Validation (equal); Visualization (equal); Writing - Original Draft (equal).
{\bf Taisei Nakamura}: Data Curation (supporting); Formal Analysis (supporting); Investigation (equal); Methodology (lead); Software (equal);  Validation (supporting); Visualization (supporting).
{\bf Masaharu Isobe}: Conceptualization (lead); Data Curation (supporting); Formal Analysis (supporting); Funding Acquisition (lead); Investigation (supporting); Methodology (equal); Project Administration (lead); Resources (lead); Software (equal); Supervision (lead); Validation (supporting); Visualization (supporting); Writing - Original Draft (lead); Writing - Review and Editing (lead).

\section*{DATA AVAILABILITY}
The source codes (Fortran 90) for local structure analysis used in this paper are openly available in the GitHub repository at  https://github.com/d-mugita/2D-LSA, which contains codes for 2D-SANNex and the NELF-A including its variants.
The data that support the findings of this study are available from the corresponding author upon reasonable request.

\end{document}